\documentclass[]{aa}

\usepackage{graphicx}
\usepackage[dvipsnames]{xcolor}
\usepackage{txfonts}
\usepackage[colorlinks=true, citecolor=blue]{hyperref}

\usepackage{multirow}
\usepackage{booktabs}
\usepackage{pdflscape}
\usepackage{placeins}
\usepackage{caption}
\usepackage{wasysym}

\newcommand{\oiii}{[O\,{\footnotesize III}]}

\newcommand{\kms}{km s$^{-1}$}

\newcommand{\myr}{$M_{\odot}$ yr$^{-1}$}
\newcommand{\msun}{$M_{\odot}$}

\newcommand{\mgas}{$M_{\rm gas}$}
\newcommand{\mdust}{$M_{\rm dust}$}
\newcommand{\mdyn}{$M_{\rm dyn}$}

\newcommand{\hi}{H\,{\footnotesize I}}

\newcommand{\lcii}{$L_{\rm [CII]}$}

\newcommand{\lir}{$L_{\rm IR}$}

\newcommand{\lsun}{$L_{\odot}$}
\newcommand{\tdust}{$T_{\rm dust}$}
\renewcommand{\arraystretch}{1.3}

\newcommand{\mgii}{Mg\,{\sc ii}}

\newcommand{\darc}{\rlap{.}{\arcsec}}
\newcommand{\cii}{[C\,{\footnotesize II}]}

\newcommand{\rubies}{RUBIES-UDS-QG-z7}

\newcommand{\lowZ}{low-$Z$}
\newcommand{\highZ}{high-$Z$}
\newcommand{\sfrir}{$\rm SFR_{IR}$}
\newcommand{\muv}{$M_{\rm UV}$}
\newcommand{\fgas}{$f_{\rm gas}$}

\usepackage{CJKutf8} %package for Chinese characters
\newcommand{\zh}[1]{\begin{CJK}{UTF8}{bsmi}#1\end{CJK}}

\begin{document} 

  \title{Extended [CII] gas emission in and around a massive quiescent galaxy at $z=7.3$}  
  \titlerunning{Extended [CII] gas emission in and around a massive quiescent galaxy at $z=7.3$}
  
  \authorrunning{F. Valentino et al.}

    \author{F. Valentino\inst{1,2}, 
            A. Pensabene\inst{1,2}, 
            A. Weibel\inst{3},
            A. de Graaff\inst{4,5}\thanks{Clay Fellow},
            D. J. Setton\inst{6}\thanks{Brinson Prize Fellow},
            P. Oesch\inst{3,1,7},
            G. Brammer\inst{1,7},
            W. M. Baker\inst{8},
            R. Bezanson\inst{9},
            J. E. Greene\inst{6},
            K. E. Heintz\inst{1,2,7},
            K. Ito\inst{1,2},
            M. Lee\inst{1,2},
            J. Leja\inst{10,11,12},
            J. Matthee\inst{13},
            B. Wang\inst{6}\thanks{NHFP Hubble Fellow},
            K. E. Whitaker\inst{14,1},
            C. C. Williams\inst{15,16},
            \and
            P. Zhu (\zh{朱芃佩})\inst{1,2,17}
            }

    \institute{
    Cosmic Dawn Center (DAWN), Denmark
    \and DTU Space, Technical University of Denmark, Elektrovej 327, DK-2800 Kgs. Lyngby, Denmark
    \and Department of Astronomy, University of Geneva, Chemin Pegasi 51, 1290 Versoix, Switzerland
    \and Center for Astrophysics, Harvard \& Smithsonian, 60 Garden St, Cambridge, MA 02138, USA
    \and Max-Planck-Institut f\"ur Astronomie, K\"onigstuhl 17, D-69117 Heidelberg, Germany
    \and Department of Astrophysical Sciences, Princeton University, 4 Ivy Lane, Princeton, NJ 08544, USA
    \and Niels Bohr Institute, University of Copenhagen, Jagtvej 128, 2200 Copenhagen N, Denmark
    \and DARK, Niels Bohr Institute, University of Copenhagen, Jagtvej 155A, DK-2200 Copenhagen, Denmark
    \and Department of Physics and Astronomy and PITTPACC, University of Pittsburgh, Pittsburgh, Pennsylvania, USA
    \and Department of Astronomy \& Astrophysics, The Pennsylvania State University, University Park, PA 16802, USA
    \and Institute for Computational \& Data Sciences, The Pennsylvania State University, University Park, PA 16802, USA
    \and Institute for Gravitation and the Cosmos, The Pennsylvania State University, University Park, PA 16802, USA
    \and Institute of Science and Technology Austria (ISTA), Am Campus 1, 3400 Klosterneuburg, Austria
    \and Department of Astronomy, University of Massachusetts, Amherst, MA 01003, USA
    \and NSF National Optical-Infrared Astronomy Research Laboratory, 950 North Cherry Avenue, Tucson, AZ 85719, USA
    \and Steward Observatory, University of Arizona, 933 North Cherry Avenue, Tucson, AZ 85721, USA
    \and INAF-Osservatorio Astrofisico di Arcetri, Largo Enrico Fermi 5, I-50125 Firenze, Italy
    }
   \date{Received --; accepted --}
    \abstract{We report the discovery of \cii\ 158\,$\mu$m emission in and around the most distant known massive quiescent galaxy \rubies\ at $z = 7.27$. Observed with ALMA in band~6, the \cii\ line independently confirms the spectroscopic redshift from JWST/NIRSpec spectra at low and medium resolution. The emission extends over an effective radius $R_{\rm eff, \,[CII]}=8 \pm 3$\,kpc, well beyond the compact stellar body traced by JWST/NIRCam ($R_{\rm eff}=209^{+33}_{-24}$ pc), with a significant fraction of $\approx 70\%$ of the flux arising from a circumgalactic halo. No dust continuum is detected at rest-frame $\sim160\,\mu$m, setting an upper limit on the infrared luminosity of $L_{\rm IR} < 1.4 \times 10^{11}\,L_\odot$, overall consistent with expectations from rest-frame UV to near-infrared SED modeling under energy balance. Converting the galaxy-scale \cii\ emission into cold gas mass, we find $\log(M_{\rm mol}/M_\odot) = 9.53^{+0.32}_{-0.31}$ and $\log(M_{\rm HI}/M_\odot) = 9.46$--$10.34$, depending on the assumed calibration and metallicity. Despite being $\approx 10\times$ more gas-poor than typical star-forming galaxies at  fixed redshift, stellar mass, and \cii\ to gas mass conversion, \rubies\ retains a substantial cold gas reservoir with fractions $f_{\rm gas} \gtrsim 20\%$ and long depletion timescales across most assumptions. The extended \cii\ halo carries approximately twice as much gas as the galaxy alone and shows a blueshifted velocity offset consistent with the tentative gas outflow detected in \mgii\ absorption in previous work, suggesting a past episode of AGN-driven gas expulsion possibly linked to the suppression of star formation. The presence of a large gas reservoir in and around a massive quiescent galaxy just 700\,Myr after the Big Bang implies that whatever mechanism is suppressing star formation must be remarkably effective at maintaining a low star formation efficiency on $\sim100$\,Myr timescales, even in the presence of abundant fuel.
    }
  \maketitle
  \nolinenumbers

\section{Introduction}  
\label{sec:introduction}
The discovery of a surprisingly large number of massive quiescent galaxies in the early universe, cemented by JWST photometric and spectroscopic observations over the last five years, poses a serious challenge to our understanding of how star formation proceeds and is rapidly suppressed in these systems (e.g., \citealt{carnall_2022, valentino_2023, nanayakkara_2024, alberts_2024, baker_2025, baker_2025_dja, merlin_2025, stevenson_2026, ji_2026_number_density_qg, zhang_2026_rubies_number_densities}). The physical mechanisms responsible for shutting down star formation remain poorly understood, and constraining the availability and properties of cold gas, the fuel for star formation, is key to distinguishing between competing quenching scenarios. It is natural to ask whether quenching is primarily driven by the complete exhaustion or expulsion of cold gas, or rather by its heating or dynamical stabilization, such that it becomes unable to collapse into new stars (see \citealt{man_2018, delucia_2025_book_quenching, whitaker-bezanson_2026} for recent reviews; e.g., \citealt{kurinchi-vendhan_2024, lagos_2024, lagos_2025, delucia_2024, xie_2024, farcy_2025, kimmig_2025, chandro-gomez_2025, chaikin_2026} for recent theoretical efforts). In the local universe, old quiescent galaxies retain low but non-negligible amounts of dust and gas, existing in both the molecular phase, traced by low-transition CO lines, and the neutral phase, observed directly via the 21\,cm \hi\ transition \citep[e.g.,][]{davis_2011, smith_2012, rowlands_2012, michalowski_2019, michalowski_2024}. Richer gas reservoirs are found in recently quenched, post-starburst systems \citep[e.g.,][]{french_2015, rowlands_2015, smercina_2018, baron2023,  ellison_2025_hi_psb}. Even so, the dense molecular phase directly associated with star formation remains elusive: the detected gas appears to reside primarily in a diffuse, low-density component traced by CO rather than the dense gas phase needed to drive collapse, suggesting that substantial amounts of gas can persist a few hundred Myr after quenching without being able to form stars \citep{french_2018, french2023}. 

At higher redshifts, powerful interferometers have pushed gas constraints on quiescent populations to cosmic noon and beyond. However, results remain mixed, depending primarily on sample selection and the choice of gas tracer, and further complicated by possible episodes of rejuvenation \citep{woodrum_2022}. Dust continuum emission has been detected in individual and stacked samples up to $z = 4.5$ \citep[e.g.,][]{gobat_2018, magdis_2021, whitaker_2021_detection, blanquez-sese_2023, Donevski+2023, Lee+2024, ji_2024}, while low- and mid-$J$ CO lines, though rarer, have been detected especially, but not exclusively, in post-starburst systems \citep{suess_2017, spilker_2018, belli_2021, williams_2021_co_quiescent, bezanson_2022, suzuki_2022, umehata_2025, lorenzon_2025, setton_2025, zanella_2023, zanella_2026, donofrio_2026}. Taken together, quiescent galaxies in the distant universe appear to retain gas fractions of $f_{\rm gas} = M_{\rm gas}/M_\star \gtrsim 1$--$10\%$, particularly shortly after quenching, possibly declining as they age and the specific star formation rate drops.  

Above $z \sim 4$, classical cold gas tracers become increasingly difficult to detect in ever smaller samples. This is where the \cii\ 158\,$\mu$m line, one of the brightest coolants of the interstellar medium (ISM), enters favorable atmospheric windows and has become the primary tracer of cold gas in star-forming galaxies and quasars (QSOs) at $z = 4-8$ \citep[e.g.,][]{lefevre_2020, bouwens_2021, herrera-camus_2025, wang_2026_aspire_survey_paper}. Building on earlier upper limits up to $z \sim 4.6$ \citep{deugenio_2023}, \cite{deugenio_2026} recently reported the first \cii\ detections in a handful of quiescent galaxies at $z = 2-3$ ($f_{\rm gas} \sim 1$--$25\%$), also investigating the connection with dust emission. Their work finds possible evidence for spatially extended gas and dust emission (see also \citealt{ji_2024}), and highlights several similarities with post-starburst systems at lower redshift, including warm dust, merger signatures, and heating from shocks, older stellar populations, and central nuclear activity.  

In this work, we push the search for \cii\ emission in quiescent galaxies to $z > 7$ and report its discovery on extended, circumgalactic scales in \rubies, the most distant massive (stellar mass $M_\star \approx 1.6\times10^{10}$ \msun) galaxy with suppressed star formation currently known \citep[$z = 7.27$;][]{weibel_2025}. Section~\ref{sec:data-red} presents the ALMA observations and data reduction, together with the available JWST ancillary data. Section~\ref{sec:results_and_discussion} presents our analysis, starting from the space of observables before converting them into physical quantities under different assumptions, whose applicability we also discuss, along with prospective tests. We summarize our conclusions in Section~\ref{sec:conclusions}. Throughout this work, we adopt a $\Lambda$CDM cosmology with $\Omega_{\rm m} = 0.3$, $\Omega_{\Lambda} = 0.7$, and $H_0 = 70\,\mathrm{km\,s^{-1}\,Mpc^{-1}}$.
% 1D spectrum
\begin{figure}
    \centering
    \includegraphics[width=\hsize]{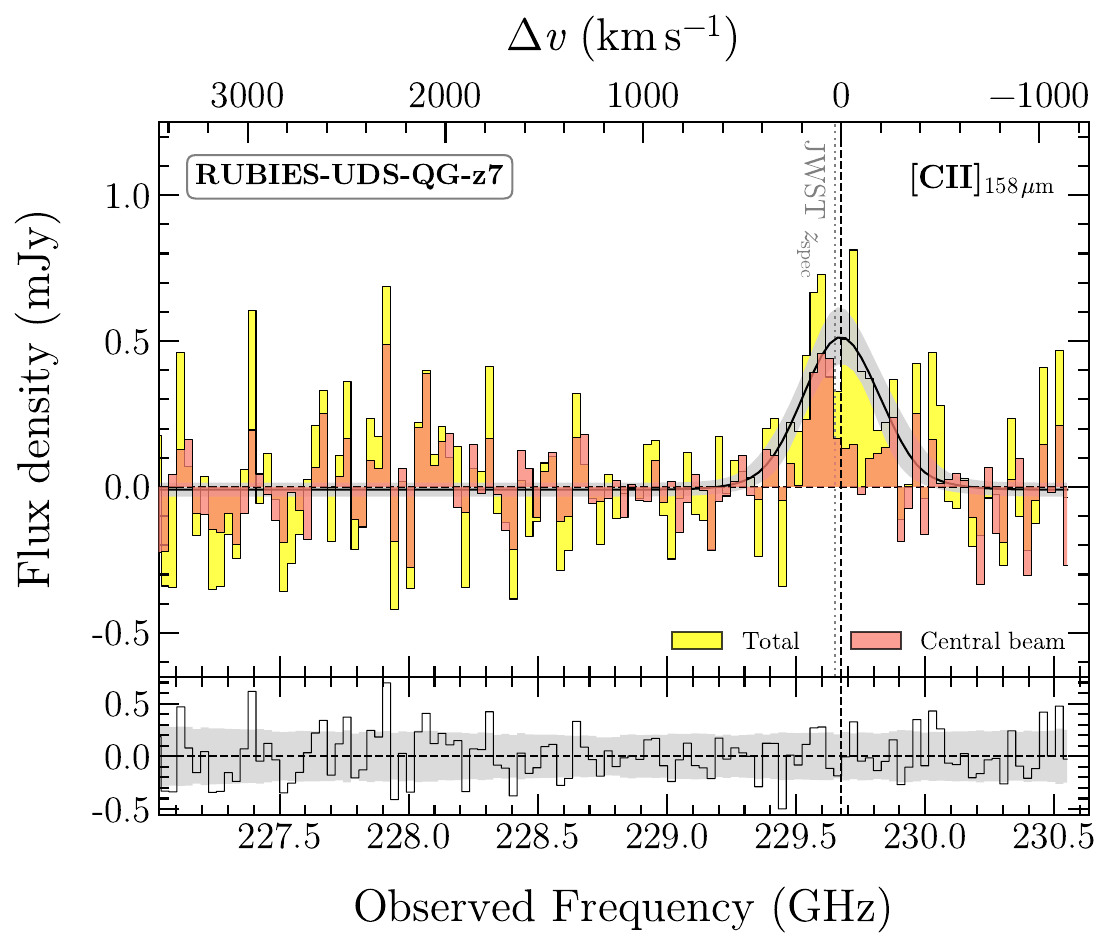}
    \caption{ALMA band 6 spectrum of \rubies. The yellow and red colors show the total emission, including the spatially extended component, and the central beam extraction that we associate with the galaxy, respectively. The black line and gray shaded area indicate the best-fit Gaussian model to the total emission and its uncertainties. The residuals are shown in the bottom inset, with the gray shaded area indicating the noise RMS. The velocity axis refers to the \cii\ line centroid of the total emission. The dotted gray vertical line indicates the redshift estimate based on JWST grating spectra \citep{valentino_2025}.}
    \label{fig:line-spec}
\end{figure}

\section{Data reduction and analysis}\label{sec:data-red}
\subsection{Description of observations}\label{ssec:descr-obs}
In this work, we make use of data collected by ALMA in Cycle 12 (program ID: 2025.1.00092.S, PI: F. Valentino) using band 6 to target the \cii\ emission line (rest-frame frequency $\nu_{\rm rest} = 1900.5369\,{\rm GHz}$) and the underlying (observed-frame) $1.3\,{\rm mm}$ dust continuum of \rubies\ ((R.A., Dec.) = ($34.4296173$, $-5.1122962$) deg) at $z=7.27$. Observations were taken within the period 17 December 2025 -- 12 May 2026 subdivided in five execution blocks (EBs). EBs were acquired in multiple array configurations (C-2 to C-5) with a number of 12-m antennas ranging from 42 to 45 and sampling baselines within $15 - 984.4\,{\rm m}$, yielding a maximum recoverable scale of $4\darc5 - 9\darc6$ and a reference Briggs-weighted synthesized beam FWHM of $0\darc48\times0\darc43$. The frequency setup consists of two pairs of 1.875-GHz wide spectral windows (SPWs) tuned for the detection of the continuum and the emission line and placed adjacently in the lower- (LSB) and upper-side band (USB), around the central frequency of 215.337 and $228.765\,{\rm GHz}$, respectively. The native spectral resolution of the observations is $\approx10.9$ and $10.2\,{\rm km\,s^{-1}}$ in the LSB and USB, respectively. The achieved noise RMS is $8\,{\rm\mu Jy}$ over the aggregate continuum-only bandwidth ($5.415\,{\rm GHz}$), and $73\,{\rm \mu Jy}$ at the representative bandwidth of $66.6\,{\rm km\,s^{-1}}$ at $229.64\,{\rm GHz}$ and reference angular resolution.

\subsection{Reduction and imaging of ALMA data}\label{ssec:red-im}
We performed standard calibration of the measurement sets (MSs) by running the \texttt{scriptForPI} delivered by the observatory, using the Common Astronomy Software Application (CASA; \citealt{McMullin+2007, Hunter+2023}) version 6.6.6.18. Self-calibration was not performed. We first inspected the data by Fourier-transforming the calibrated data using the CASA task {\tt tclean} with a natural-weighting scheme of the visibilities. We obtained dirty datacubes of the LSB and USB separately and an aggregate LSB+USB continuum map. We set the pixel size to $0\darc09$ to achieve the Nyquist-sampling of the synthesized beam size, and the channel size to $40\,{\rm km\,s^{-1}}$ for the datacube. The \cii\ line emission of \rubies\ is detected in the USB cube around the observed-frame frequency of $229.7\,{\rm GHz}$, which is consistent with the expectation based on the spectroscopic redshift derived from the JWST/NIRSpec spectra (Section \ref{ssec:ad}). The line emission appears extended at the resolution of the data ($0\darc64\times0\darc55$). No emission is detected in the continuum map at the location of the galaxy, implying a $3\sigma$ upper limit on the $1.3\,{\rm mm}$ flux density of $F_{\rm 1.3\,{\rm mm}}(3\sigma)<18\,{\rm \mu Jy}$. 
% JWST+contours line velocity split + PV
\begin{figure*}
    \centering
    \includegraphics[width=\textwidth]{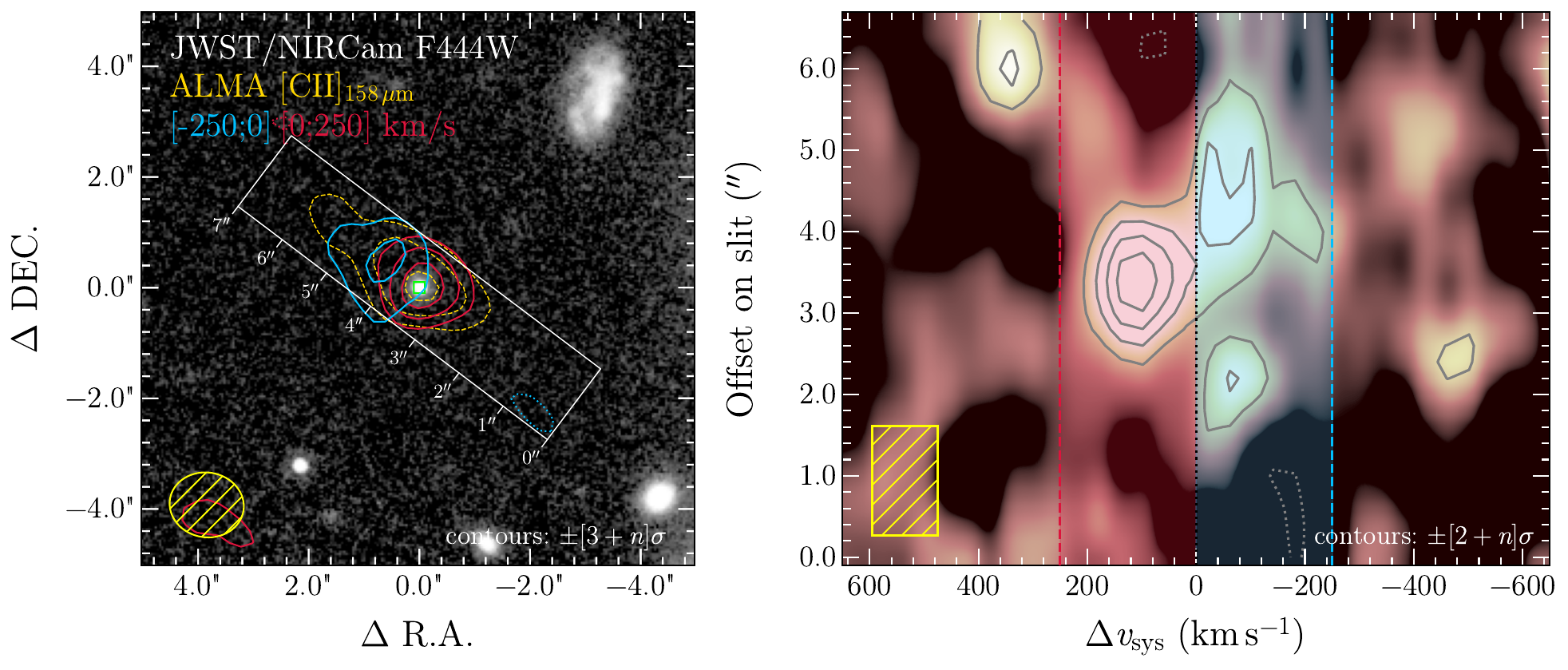}
    \caption{Extended \cii\ emission around \rubies. Left panel: the \cii\ emission is indicated by sets of contours: thick dashed yellow lines indicate the total moment-0 map within $\pm2\sigma\approx\pm390\,{\rm km\,s^{-1}}$ around the total \cii\ line centroid, while solid red and blue lines indicate the emission redward ($[0, 250]\,{\rm km\,s^{-1}}$) and blueward ($[-250, 0]\,{\rm km\,s^{-1}}$) of the systemic redshift. The contours are spaced as multiples of the noise level, starting at $\pm3\sigma$. Negative peaks are marked by dotted lines with the same color scheme. The beam size is shown in the bottom left corner of the cutout. The background image is from JWST/NIRCam with the F444W filter and the green square indicates the center of the stellar emission. The white box marks the pseudoslit used in the position-velocity diagram in the right panel. Right panel: position-velocity diagram. The gray contours are at $\pm2\sigma$ significance. The image was smoothed along the velocity axis using a uniform filter of width equal to $120\,{\rm km\,s^{-1}}$ (i.e., three channels of the datacube as shown in the bottom left corner). Velocity ranges for the red and blue map shown in the left panel are also highlighted.}
    \label{fig:map_pv}
\end{figure*}

We therefore obtained three sets of cleaned datacubes of the USB by applying different tapering schemes of the baselines to optimize the recovery of the extended line emission. We ran {\tt tclean} down to $1.5\sigma$ per channel within 3D circular masks of $2\arcsec$-radius. We adopted a natural weighting for the visibilities to maximize the surface brightness sensitivity and {\tt MultiScale CLEAN} \citep{Cornwell2008} as deconvolution algorithm with {\tt scales=[0,5,15]} to optimize the modeling of the emission from compact point-like to moderately extended sources (up to $\sim 3\times$ the beam size). In addition to the standard weighting scheme, we applied a Gaussian tapering in the $u\varv$ plane with ${\rm FWHM} = 0.5\arcsec$ and $0.8\arcsec$ (corresponding to ${\rm HWHM}_{u\varv} \approx 182$ and $114\,{\rm k\lambda}$), thereby downweighting the longest baselines and yielding a synthesized beam FWHM of $0\darc98\times0\darc88$ and $1\darc33\times1\darc16$, respectively. The non-tapered and tapered datacubes were obtained by setting pixel scales of $0\darc09$ and $0\darc2$, respectively, and a channel width of $40\,{\rm km\,s^{-1}}$. The resulting noise RMS is $0.09$, $0.11$, and $0.13\,{\rm mJy\,beam^{-1}\,channel^{-1}}$ for the untapered, $0\darc5$-, and $0\darc8$-tapered datacubes, respectively. 

\subsection{Line flux and luminosity measurement}
\label{ssec:line-flux}
We measured the \cii\ line flux of \rubies\ by extracting the galaxy emission line obtained through the $2\sigma$-clipped photometry method \citep[e.g.,][]{Bethermin+2020}. This method enables the recovery of the line flux with an optimal S/N while preventing significant flux losses. In brief: 1) we performed a single-pixel extraction of the spectrum at the source location\footnote{We adopted the peak of the stellar emission in the NIRCam/F444W image as the source location (Figure~\ref{fig:map_pv}).} 
and fit the line with a Gaussian plus a constant for the continuum; 2) we obtained a moment-0 map by collapsing channels within $\pm2\sigma$ of the line centroid; 3) we extracted a new spectrum from the cube by integrating the signal from all pixels above ${\rm S/N\geq2}$ and we fit the new line profile. We repeated steps 2) and 3) until convergence. This analysis showed that the source flux recovery is maximized in the $0\darc8$-tapered cube and we therefore adopted it as our reference dataset. For the rest of this work, we assume that the single-pixel spectral extraction at the peak of the stellar light from step 1) (i.e., the "central beam") is associated with the \cii\ line emission of the galaxy. However, we note that, at the resolution of the data (${\rm FWHM}\approx3.1-6.3\,{\rm kpc}$ depending on the tapering scheme), this is likely an upper limit on the true galaxy flux, as we may be including some emission from more extended scales. 

Finally, we performed a fit of the total line profile by minimizing the residuals between the data and a Gaussian+constant model using the MCMC ensemble sampler {\tt emcee} Python package \citep{Foreman+2013} with 100 walkers and 3000 steps, discarding the first 1000 as a burn-in phase. We adopted Gaussian uncertainties in the definition of the likelihood and flat priors on the free parameters based on the last fitting iteration as described above. We derived the line luminosity as \citep{Solomon+1997}:
$L_{\rm [CII]}\,[L_{\astrosun}]=1.04\times10^{-3}\,F_{\rm [CII]}\,\nu_{\rm obs}\,D_{\rm L}^{2}$, $
L^\prime_{\rm [CII]}\,[{\rm K\,km\,s^{-1}pc^{2}}]=3.25\times10^{7}\,F_{\rm [CII]}\,{D_{\rm L}^2\,}{(1+z)^{-3}\,\nu_{\rm obs}^{-2}}$,
where $F_{\rm [CII]}$ is the velocity-integrated line flux in units of ${\rm Jy\, km\,s^{-1}}$, $\nu_{\rm obs}=\nu_{\rm rest}/(1+z)$ is the observed central frequency of the line in GHz at the redshift $z$, and $D_{\rm L}(z)$ is the luminosity distance in Mpc. The best-fit line centroid corresponds to a source redshift of $z_{\rm \cii}=7.2749^{+0.0011}_{-0.0012}$ for the total line flux extraction and $z_{\rm \cii}=7.2774^{+0.0005}_{-0.0006}$ for the central beam. These are overall consistent with the spectroscopic redshifts from the PRISM ($z=7.287^{+0.007}_{-0.006}-7.290^{+0.005}_{-0.006}$ in the range of plausible SED models in \citealt{weibel_2025}, Section \ref{ssec:ad}) and grating spectra ($z=7.2758 \pm 0.0011$, \citealt{valentino_2025}).

In Table~\ref{tab:line-flux}, we report the measurements and derived quantities from the modeling of the \cii\ emission line. In Figure~\ref{fig:line-spec}, we show the source spectrum extracted from the central beam and integrated over the total line emission, along with the best-fit model.

% Extended emission
\subsection{Modeling of the extended \cii\ emission}\label{ssec:mod-em}
The \cii\ line emission from \rubies\ is spatially
resolved. This is evident from the extraction of line fluxes with only $\approx30\%$ of the total line luminosity enclosed in the central beam (Table~\ref{tab:line-flux}). In Figure~\ref{fig:map_pv}, we inspect the kinematics of the extended \cii\ by superimposing pseudo-narrow-band images obtained by integrating channels in the datacube blueward ($[-250,0]\,{\rm km\,s^{-1}}$) and redward ($[0,250]\,{\rm km\,s^{-1}}$) of the total line centroid. The peak emission of the blue and red components appears significantly offset, suggesting the presence of a velocity gradient along the South-West--North-East direction. The red component is associated with the bulk of the line emission and centered around the peak of the stellar emission. This conclusion is further supported by the analysis of the position-velocity (PV) diagram extracted from a pseudo-slit aligned with the observed velocity gradient. 

We estimated the source size by modeling its morphology directly in the visibility $(u,v)$ plane. Given the relatively low S/N of the observations, visibility-domain fitting is preferable to image-plane analyses because it avoids potential biases introduced by the deconvolution and reconstruction of interferometric images from incompletely sampled $u\varv$-data. This is especially relevant in our case, as the observations were acquired in multiple array configurations, resulting in sparse $u\varv$-coverage. We combined the calibrated MSs into a single one and split target visibilities corresponding to 20 channels of $40\,{\rm km\,s^{-1}}$ width within $\pm390\,{\rm km\,s^{-1}}$ of the total line systemic redshift using the {\tt concat} and {\tt mstransform} tasks of CASA, respectively.  We fit the visibilities in the $(u,v)$ plane using UVMODELFIT \citep{Marti-Vidal+2014}; given the low per-channel S/N, we fit the moment-0 map of the line emission rather than the full 3D datacube ($u,v,\nu$).
% Table ALMA
\begin{table}[t]
\def\arraystretch{1.15}
\caption{\cii\ line and continuum flux measurements, luminosity estimates, and derived quantities of \rubies.}  
\label{tab:line-flux}    
\centering
\resizebox{\hsize}{!}{
\begin{tabular}{l c c}
\toprule
\toprule
Line property  & Total & Central beam$^{(\dagger)}$\\
\cmidrule(lr){1-3}
Redshift & $7.2749^{+0.0011}_{-0.0012}$ & $7.2774^{+0.0005}_{-0.0006}$ \\
FWHM $[{\rm km\,s^{-1}}]$ & $465^{+129}_{-103}$ & $189^{+112}_{-49}$ \\
Flux $[{\rm Jy\,km\,s^{-1}}]$ & $0.26^{+0.06}_{-0.05}$ &  $0.08^{+0.02}_{-0.02}$ \\
$L_{\rm [CII]}$ $[10^8\,L_{\astrosun}]$ & $3.4^{+0.8}_{-0.7}$ & $1.1^{+0.3}_{-0.2}$ \\
$L'_{\rm [CII]}$ $[10^{10}\,{\rm K\,km\,s^{-1}\,pc^2}]$ & $0.16^{+0.04}_{-0.03}$ & $0.05^{+0.01}_{-0.01}$ \\
\midrule
\multicolumn{3}{c}{Observed-frame 1.3-mm continuum}  \\
\cmidrule(lr){1-3}
$S_{\rm 1.3\,mm}$ $[{\rm \mu Jy}]$ & \multicolumn{2}{c}{$< 18$}\\
$L_{\rm IR,\,8-1000\,\mu m}$ [$10^{11}$ \lsun]& \multicolumn{2}{c}{$< 1.4$}\\
$M_{\rm dust}$ [$10^{6}$ \msun]& \multicolumn{2}{c}{$< 6.3$}\\
\bottomrule
\end{tabular}
}
\tablefoot{All quantities are measured from the $0\darc8$-tapered datacube. $^{(\dagger)}$Quantities measured from the spectrum extracted from the peak of the stellar emission in the JWST/F444W image. \lir\ and \mdust\ are derived at fixed $T_{\rm dust}=40\,\mathrm{K}$ and $\beta_{\rm IR} = 1.8$ (Section \ref{ssec:dust}).}
\end{table}

\begin{figure}
    \centering
    \includegraphics[width=\hsize]{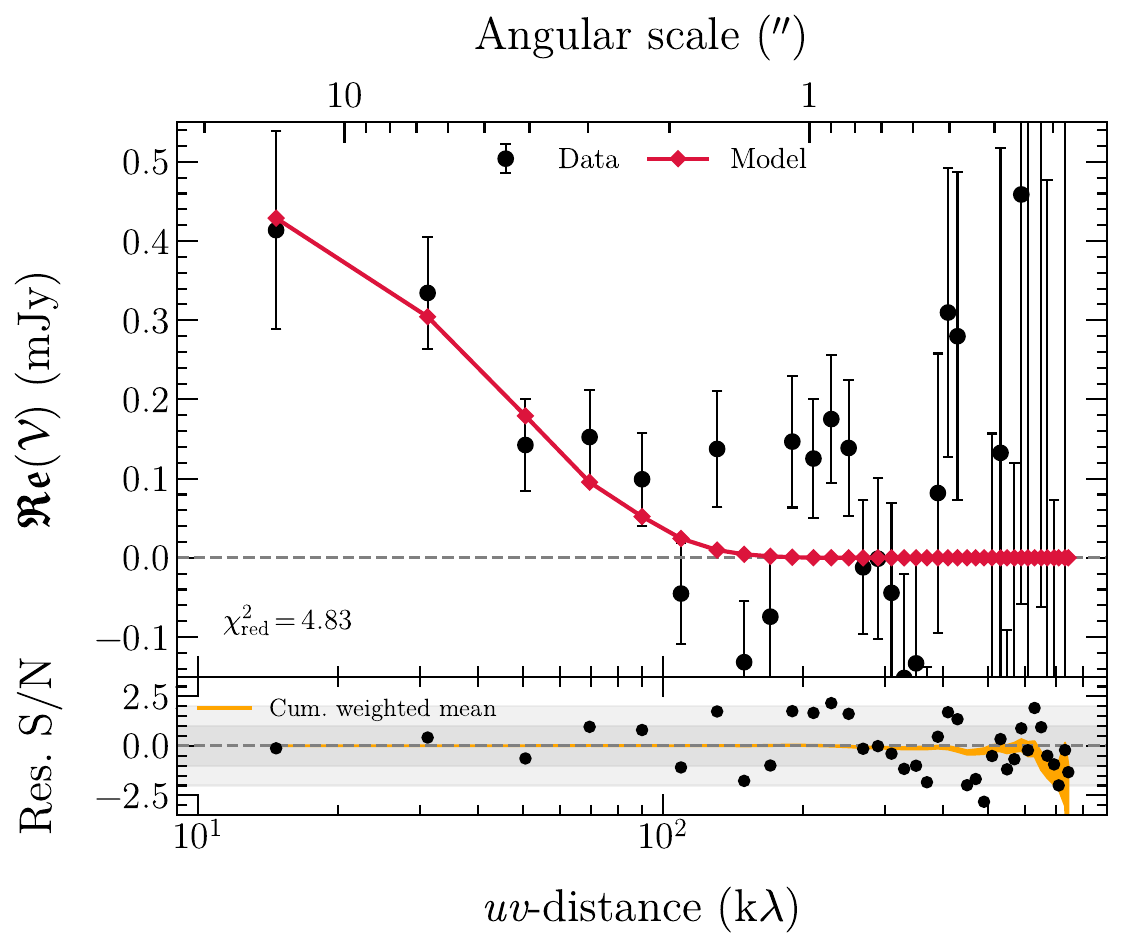}
    \caption{Morphological modeling of \rubies\ in the visibility domain. {Top panel:} The real part of the visibilities (black points) binned to $20\,{\rm k}\lambda$ ($\lambda=1.305\,{\rm mm}$). The best-fit Gaussian model is shown in red. The reduced chi-squared value is reported in the bottom left corner. {Bottom panel:} Residual S/N after subtracting the best-fit model ([data$-$model]/error; black points). The orange curve shows the cumulative variance-weighted mean of the residuals. Dark and light gray shaded areas indicate $1-$ and $2\sigma$ levels. 
    }
    \label{fig:uv_plot}
\end{figure}

\begin{figure*}
    \centering
    \includegraphics[width=\hsize]{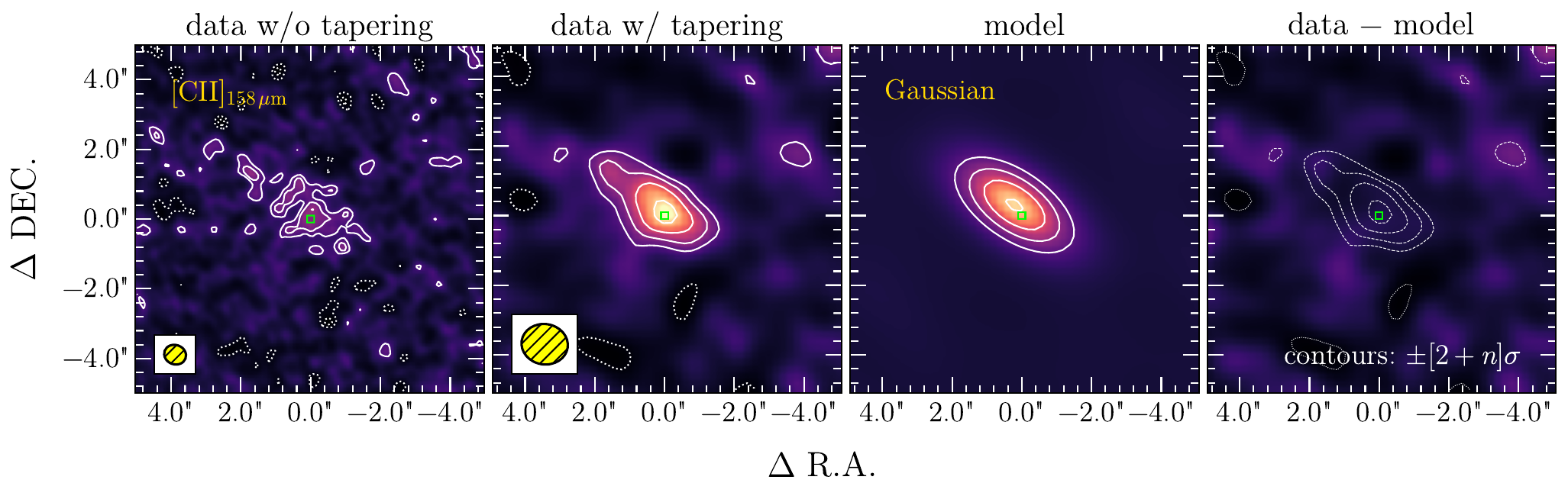}    
    \caption{Maps of the \cii\ emission and best-fit model of \rubies\ in the image plane. From left to right: untapered and tapered (${\rm HWHM}_{uv} \approx 114\,{\rm k\lambda}$) moment-0 dirty maps ($\pm390\,{\rm km\,s^{-1}}$), the best-fit Gaussian model, and residuals with data contours superimposed. White contours indicate $\pm[2+n]\sigma$ levels, where $\sigma$ is the noise RMS and $n \geq 0$ is an integer. The synthesized beam FWHM is shown in the bottom left corner. The green square marks the peak of the stellar emission.} 
\label{fig:morph-mod}
\end{figure*}

We employed a single 2D Gaussian component to model the source emission. There are six free parameters: the coordinates of the center, the major axis FWHM, the axis ratio, the position angle, and the total enclosed flux. We adopted flat priors and minimized the residuals between the data and the model using standard $\chi^2$ statistics. Finally, we extracted the data and best-fit model visibilities using the {\tt export\_uv\_table} task of the UVPLOT Python package \citep{Tazzari2017}. In Figure~\ref{fig:uv_plot}, we report the binned visibilities with the best-fit model. In Figure~\ref{fig:morph-mod}, we show the results in the image plane after a Fourier transform of the data and the model using the {\tt tclean} task in CASA. The results suggest that the model adequately reproduces the data, with residuals consistent with the noise RMS at the $\lesssim2\sigma$ level. In particular, the Gaussian model reproduces well the extended structure of the source, which appears at $>1-2\arcsec$ angular scales. The best-fit model returns an effective radius of $R_{\rm eff}={\rm FWHM}/2=1.7\pm0.6\,{\rm arcsec}$, or $8\pm3\,{\rm kpc}$ at the redshift of the galaxy.

% UV modeling results

% Morph. modeling results

% Optical and near-infrared imaging and spectroscopy: Weibel+25, Valentino+25
\subsection{Optical and near-infrared photometry and spectroscopy}\label{ssec:ad}
\rubies\ benefits from a wealth of JWST ancillary data, including NIRCam and MIRI imaging and NIRSpec spectroscopy. In this work, we make use of the data and modeling described in \cite{weibel_2025} and \cite{valentino_2025}. Briefly, \cite{weibel_2025} modeled the $0.9$--$18\,\mu$m JWST photometry collected as part of the Public Release IMaging for Extragalactic Research (PRIMER) program (GO-1837, PI: J.~Dunlop; \citealt{donnan_2024}) and the PRISM spectrum from the Red Unknowns: Bright Infrared Extragalactic Survey (RUBIES, GO-4233, PIs: A.~de Graaff and G.~Brammer; \citealt{degraaff_2024_rubies}) with {\sc Prospector} \citep{johnson_2021_prospector}. The models assume the Flexible Stellar Population Synthesis package \citep[FSPS;][]{conroy_2010} with the MILES spectral library \citep{sanchez-blazquez_2006}, MIST isochrones \citep{choi_2016}, a \cite{chabrier_2003} initial mass function, the \cite{kriek_2013} dust law, and a non-parametric star formation history (SFH) with a continuity prior \citep{leja_2019_sfh}. The SFH is split into 8 time bins of 10, 40, and 50 Myr for the most recent 100 Myr of cosmic time, followed by 5 bins of approximately $125$ Myr each. The images (v7) and spectra (v4) are available through the Dawn JWST Archive (DJA\footnote{\url{https://dawn-cph.github.io/dja/}}; \citealt{degraaff_2024_rubies, heintz_2025, valentino_2023, valentino_2025}).

\cite{weibel_2025} identifies two sets of best-fit parameters converging on a low ($Z=0.11^{+0.02}_{-0.01}\,Z_\odot$) and a high ($Z=1.19^{+0.24}_{-0.27}\,Z_\odot$) stellar metallicity solution, reflecting unresolved degeneracies and highlighting the limitations of current models \citep{deGraaff_2025}, which have until recently lacked consistent $\alpha$-enhanced isochrones and spectra \citep{park_2025_alphaMC}.
The true stellar metallicity of \rubies\ plausibly lies between these two extremes.
Both solutions return consistent redshifts ($z = 7.287-7.290$), stellar masses ($\log(M_\star/M_\odot) \approx 10.2$), and overall SFH shapes, with an abrupt decline in SFR over the last 100 Myr and consistently low SFR estimates of $\approx 1$ \myr\ averaged over 10 or 50 Myr. These values are also consistent with the SFR inferred from upper limits on the rest-frame optical emission lines, probing similar timescales ($\rm SFR_{Lines} \lesssim 6$ \myr). The SFR averaged over 100 Myr is formally consistent between the two solutions given the large uncertainties, though the nominal values differ substantially ($\rm SFR_{100} = 0.84^{+20.16}_{-0.78}$ and $48.89^{+21.12}_{-13.04}$ \myr for the low- and high-$Z$ solutions, respectively).

Medium-resolution spectra from RUBIES (G395M/F290LP) and the DeepDive program (G235M/F170LP; GO-3567, PI: F.~Valentino; \citealt{ito_2025_deepdive}) were combined to yield a refined redshift of $z = 7.2758 \pm 0.0011$ \citep{valentino_2025}.  This value is consistent with that from the \cii\ emission (Figure \ref{fig:line-spec}), unambiguously confirming the spectroscopic redshift of the galaxy. Independent SED modeling of the G235M spectrum and photometry in \cite{valentino_2025} returns results broadly consistent with those in \cite{weibel_2025}. 

Throughout the remainder of this analysis, we adopt the modeling from \cite{weibel_2025}, explicitly quoting parameter ranges spanned by the low- and \highZ\ solutions where relevant. For convenience, we report the key parameters in Table \ref{tab:physical_properties}.

% Results and discussion
\section{Results and discussion}
\label{sec:results_and_discussion}

% Dust emission
\subsection{Dust emission}
\label{ssec:dust}
We converted the upper limit on the dust continuum emission at rest-frame $\sim160\,\mu$m into an upper limit on the total infrared luminosity, $L_{\rm IR,\,8-1000\,\mu m} < 1.4\times10^{11}\,L_\odot$, using a single modified blackbody with fixed dust temperature $T_{\rm dust} = 40\,\mathrm{K}$ and emissivity index $\beta_{\rm IR} = 1.8$, which we modeled with \textsc{Mercurius} \citep{witstok_2023} and accounting for the effect of the cosmic microwave background \citep[CMB;][]{dacunha_2013}. These values of \tdust\ and $\beta_{\rm IR}$ are consistent with those typical of star-forming galaxies at $z = 7$ \citep{witstok_2023}. The lower \tdust\ reported for the dusty interstellar medium of lower-redshift quiescent galaxies \citep[$T_{\rm dust} \approx 20\,\mathrm{K}$;][]{magdis_2021} is not physical at this redshift given the temperature floor imposed by the CMB ($T_{\rm CMB}=22.6$ K). The upper limit on \lir\ is overall consistent with the posterior distributions from the SED modeling with \textsc{Prospector} under the assumption of energy balance ($\mathrm{log}(L^{\rm pred}_{\rm IR}/L_\odot)=10.89^{+0.12}_{-0.12}$ and $11.05^{+0.11}_{-0.11}$ for the high- and \lowZ\ solutions, respectively). Although insufficient to tightly constrain the broadband SED models at this stage, the dust continuum upper limit hints at mild tension with the \lowZ\ solution, suggesting that future dust observations at higher frequencies could provide actual \lir\ constraints and help break the degeneracies in the models (Section \ref{ssec:caveats-tests}). Using the frequency-dependent dust mass absorption coefficient $\kappa = \kappa_0(\nu/\nu_0)^{\beta_{\rm IR}}$, with $\kappa_0 = 8.94\,\mathrm{cm^2\,g^{-1}}$ at rest-frame $158\,\mu$m \citep{hirashita_2014}, we derived an upper limit on the dust mass of $M_{\rm dust} < 6.3\times10^{6}\,M_\odot$ ($3\sigma$). For reference, applying the conversion of \cite{kennicutt-evans_2012} for star-forming galaxies, the limit on \lir\ formally corresponds to $\rm SFR_{\rm IR} < 21\,$\myr, though this conversion is known to significantly overestimate the true SFR in recently quenched galaxies primarily due to dust heating from A-type stars \citep{utomo_2014, leja_2019_older_quiescent_universe}. We therefore do not use this estimate in the rest of the analysis. We further caution that the upper limits on \lir\ and \mdust\ depend sensitively on the assumed modified blackbody parameters and should be interpreted accordingly.

% Observed parameter space
\subsection{The space of observables}
\label{ssec:observables}
To place the \cii\ detection in context, we compare \rubies\ with star-forming galaxies at $z = 4-8$ from the ALMA Large Program to INvestigate \cii\ at the Early times \citep[ALPINE;][]{lefevre_2020, Bethermin+2020, Faisst+2020} and the Reionization Era Bright Emission Line Survey \citep[REBELS;][]{bouwens_2021}; low-luminosity quasars at $z \sim 6$ from the Subaru High-z Exploration of Low-luminosity Quasars \citep[SHELLQs;][]{Izumi+2018, Izumi+2019} project, and quiescent galaxies at $z = 2-4.6$ from \cite{deugenio_2026}.
We focus here on purely observed quantities and defer
their physical interpretation to the following Sections.

\begin{figure}
    \centering
    \includegraphics[width=\columnwidth]{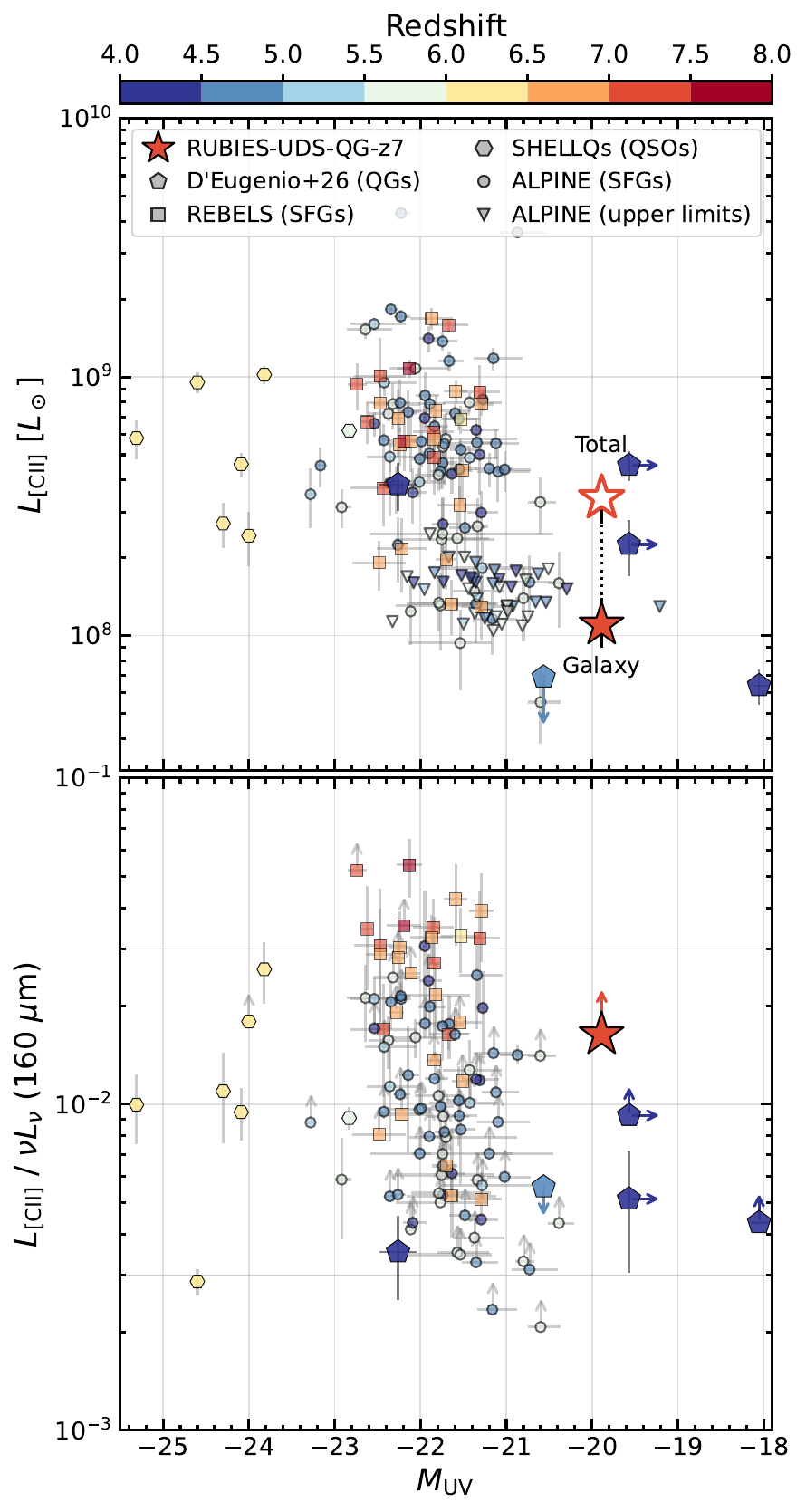}
    \caption{\cii, dust, and UV emission of high-redshift galaxies. Top panel: \cii\ luminosity as a function of rest-frame UV magnitude. Bottom panel: ratio of \cii\ luminosity to the underlying dust continuum at $160\,\mu$m rest-frame, as a function of rest-frame UV magnitude. In both panels, the filled and empty stars indicate the \cii\ emission associated with the \rubies\ galaxy and the total emission including the extended halo, respectively; pentagons the QGs from \cite{deugenio_2026}, circles and squares the SFGs from ALPINE and REBELS, hexagons the low-luminosity QSOs from SHELLQs, and downward triangles upper limits from ALPINE. Symbols are color-coded by redshift.}
    \label{fig:observed_properties}
\end{figure}

Figure~\ref{fig:observed_properties} shows \lcii\ and its
ratio with the dust continuum at $160\,\mu$m rest-frame,
both as a function of rest-frame UV magnitude at
$1500\,\AA$\footnote{\muv\ is evaluated at $1800\,\AA$
rest-frame for the \cii-detected MRG-M0138 quiescent galaxy in \cite{deugenio_2026}, based
on an HST $V$-band detection from \cite{akhshik_2023}.
Values are not corrected for lensing.}.
Star-forming galaxies, QSOs, and most quiescent galaxies occupy distinct regions of this parameter space. Star-forming galaxies follow a \lcii--\muv\ correlation across the full redshift range, reflecting the unobscured
SFR--\lcii\ relation. 
For QSOs, the nuclear emission brightens \muv\ without
significantly contributing to the \cii\ emission of the host galaxy.
Among the quiescent galaxies, the lensed galaxy MRG-M1038 at $z = 1.98$ \citep{newman_2018, newman_2025, akhshik_2023} lies on
the correlation for star-forming galaxies, while QG1--3 at $z = 3.09$
\citep{kubo_2016, kubo_2021, kubo_2022,
umehata_2025, umehata_2026} are outliers at faint \muv.
The $z = 4.6$ system from \cite{carnall_2023b}, undetected in \cii, 
appears consistent with an extrapolation of the correlation toward 
the faint end. Considering only the \cii\ emission associated with the
galaxy, \rubies\ falls in the region of parameter space occupied by the \muv-faint, \cii-detected quiescent galaxies in \cite{deugenio_2026}.
Including the extended \cii\ halo places \rubies\ well
outside the correlation for star-forming galaxies, at faint \muv\ and high \lcii. The \lcii-to-continuum ratio in the bottom panel of Figure 
\ref{fig:observed_properties} tells a similar story: even without 
the halo, the non-detection of the continuum already pushes \rubies\ 
outside the correlation established for star-forming galaxies at 
lower and similar redshift, into the region occupied by other 
quiescent galaxies, though with a more extreme and better 
constrained ratio thanks to the clear \cii\ detection and stringent 
upper limit on the dust continuum.

% SFR - LCII
\begin{figure}
    \centering
    \includegraphics[width=\columnwidth]{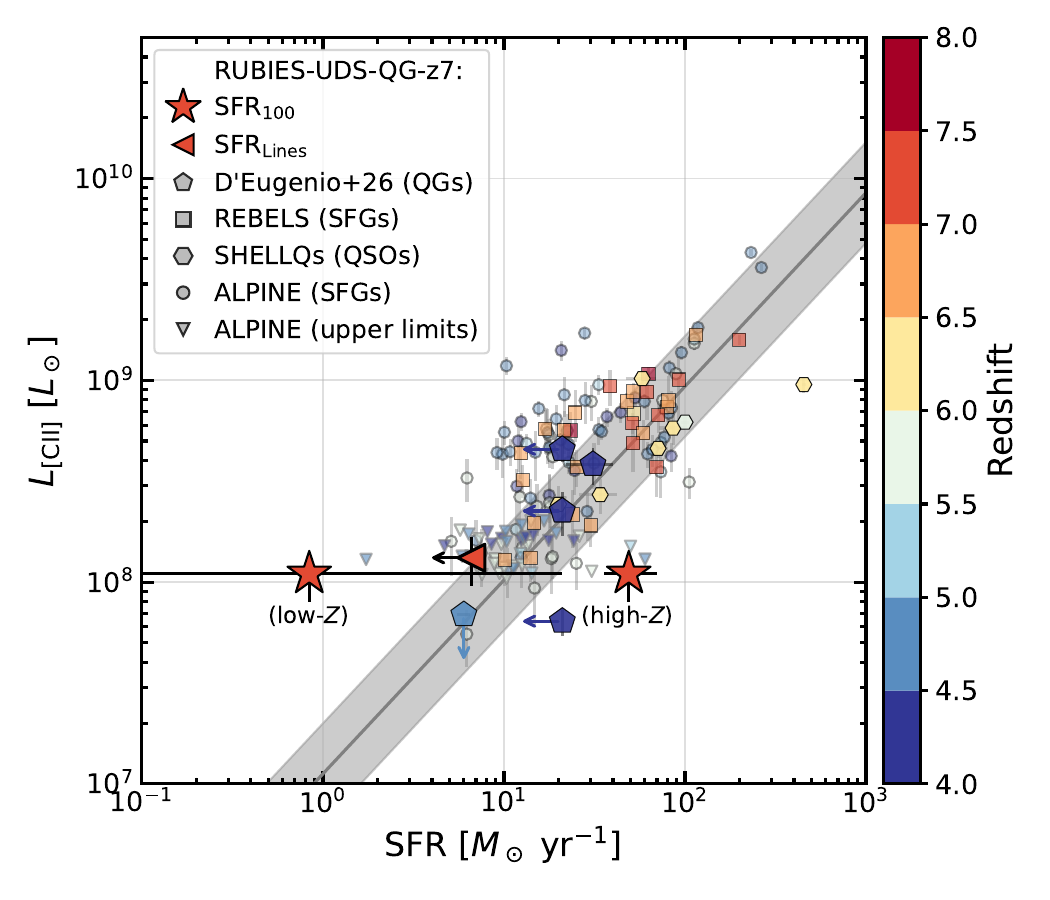}
    \caption{\lcii--total SFR relation. The gray line and shaded area show the \lcii--SFR relation and its scatter from \cite{schaerer_2020}. Large filled symbols indicate the SFR estimates for \rubies: stars mark the values for $\rm SFR_{100}$ from the low- and \highZ\ solutions of the spectrophotometric modeling in \cite{weibel_2025}; the leftward triangle shows the upper limit from rest-frame optical lines emission. \lcii\ refers to the galaxy component only, excluding the extended emission. Remaining symbols are as in Figure~\ref{fig:observed_properties}, color-coded by redshift.}
    \label{fig:sfr_lcii}
\end{figure}

\subsection{The \lcii--SFR relation}  
\label{ssec:lcii-sfr}
The observables in Figure~\ref{fig:observed_properties} can be mapped onto physical properties under different assumptions. Figure~\ref{fig:sfr_lcii} shows the classical \lcii--total SFR relation established both locally and at high redshift \citep[e.g.,][]{delooze_2014, schaerer_2020}. Using the \cii\ emission from the central beam that we associate with the galaxy, \rubies\ is broadly consistent with the \lcii--SFR relation when considering SED solutions averaged over 100 Myr. However, the measured \lcii\ luminosity is higher than that predicted by the relation assuming SFRs inferred on timescales of 10 Myr from the SED modeling ($\rm SFR_{10}=0.64^{+0.83}_{-0.60}$ and $1.08^{+1.55}_{-0.98}$ \myr\ for the low- and \highZ\ solutions, respectively; Table \ref{tab:physical_properties}) and the upper limit from the emission lines, suggesting that the recent drop in SFR is not accompanied by a complete removal of the cold gas reservoir. Star-forming galaxies from ALPINE and REBELS follow a clear correlation, largely reflecting that between \lcii\ and \muv. The correspondence is not strictly one-to-one, as we show the total $\rm SFR = SFR_{UV} + SFR_{IR}$ where an estimate of \sfrir\ is available from far-infrared detections. Low-luminosity quasars also broadly follow the relation once the nuclear contribution to \muv\ is removed. Quiescent galaxies with \cii\ detections from \cite{deugenio_2026} are broadly consistent with the star-forming galaxy locus, though to varying degrees. For MRG-M0138, the SFR from optical and near-IR modeling ($\rm SFR = 31 \pm 9$ \myr; \citealt{newman_2018}) is fully consistent with the \lcii--SFR relation of \cite{schaerer_2020} and with \sfrir, as also noted by \cite{deugenio_2026}. For QG1--3 in SSA22, the upper limit on \sfrir\ from \cite{kubo_2021} is formally consistent with the relation, while SFR values from SED modeling and optical emission lines are substantially lower; a similar situation holds for the $z = 4.6$ quiescent galaxy of \cite{carnall_2023b}. As discussed in \cite{deugenio_2026} and in Section \ref{ssec:dust}, \sfrir\ based on conversions valid for star-forming galaxies tends to overestimate the SFR in quiescent galaxies relative to SED- and line-based values due to dust heating by sources other than young UV-bright stars, such as older stellar populations \citep[particularly relevant for recently quenched systems, e.g.,][]{belli_2021}, a central AGN \citep{ji_2024}, and turbulence or shocks from gravitational interactions with nearby galaxies.

\subsection{Substantial gas reservoirs after quenching}  
\label{ssec:gas_properties}
To quantify the gas content of \rubies, we estimated \mgas\ under different assumptions. \cii\ emission arises from the ionized, neutral, and molecular gas phases, though the bulk is generally associated with the colder and denser phases \citep[e.g.,][]{vallini_2015, madden_2020, vizgan_2022_cii_molecular, vizgan_2022_cii_neutral, wolfire_2022, fudamoto_2025}. We adopted two calibrations: the empirical conversion to the molecular gas mass from \cite{zanella_2018}, $M_{\rm mol} = 31\,L_{\rm [CII]}$ (0.3 dex scatter), calibrated for main-sequence and starburst galaxies over a wide redshift range, which also allows direct comparison with \cite{deugenio_2026}; and the metallicity-dependent \cii-to-\hi\ conversion from \cite{heintz_2021}, $\log(M_{\rm HI}/M_\odot) = (-0.87\pm0.09)\,\log(Z/Z_\odot) + (1.48\pm0.12) + \log(L_{\rm [CII]}/L_\odot)$, calibrated against distant, mostly low-metallicity gamma-ray bursts. Considering only the galaxy \cii\ emission from the central beam for consistency with SFR estimates from JWST imaging and spectroscopy, we find $\log(M_{\rm mol}/M_\odot) = 9.53^{+0.32}_{-0.31}$ (independent of metallicity), and $\log(M_{\rm HI}/M_\odot) = 10.34^{+0.19}_{-0.17}$ and $9.46^{+0.18}_{-0.17}$ for the low- and \highZ\ solutions, assuming $Z_{\rm gas} = Z_\star$ in the absence of a direct gas-phase metallicity measurement. 

Converting the upper limit on \mdust\ via the gas-to-dust ratio ($\delta_{\rm GDR}$) of \cite{popping_2023}, taking its relation with metallicity to be unchanged with redshift (\citealt{popping_2022, heintz_2023_dtg}, but see \citealt{heintz_2025_inefficient_dust_production}) and accounting for the strong assumptions in Section \ref{ssec:dust}, yields $\log(M_{\rm gas}/M_\odot) < 10.44$ and $< 9.13$ for the low- and \highZ\ solutions, in general agreement with the \hi\ and molecular mass estimates. The gas-to-dust ratio for the \highZ\ solution is about twice as high as the common value of $\delta_{\rm GDR}=100$ \citep{de-vis_2019}. Adopting $\delta_{\rm GDR}=100$ would thus return a gas mass of $\log(M_{\rm gas}/M_\odot)<8.8$. Reversing the argument, assuming that \cii\ traces the molecular gas in \rubies, we derive a molecular gas-to-dust ratio $\delta_{\rm GDR}\gtrsim540$. Such a high value would be consistent with recent observational and theoretical results suggesting that dust is destroyed on shorter timescales than those on which the gas is expelled or consumed after quenching \citep{whitaker_2021,lorenzon_2025,spilker_2025}.

The \mgas\ estimates are broadly consistent with an estimate of the dynamical mass from the \cii\ emission based on the virial theorem, even when accounting for the stellar mass ($M_\star\sim1.6\times10^{10}\,M_\odot$): $M_{\rm dyn} = 5\sigma_{\rm vel}^2 R/G = 2.3^{+3.5}_{-1.0} \times 10^{10}\,M_\odot$, where $\sigma_{\rm vel} = \rm FWHM_{\rm [CII]} / (2\sqrt{2\ln 2})$ is the velocity dispersion of the \cii\ line, $R$ is the size of the emission associated with the galaxy, and $G$ is the gravitational constant. We adopted $\rm FWHM_{\rm [CII]} = 189^{+112}_{-49}$ \kms\ from the central beam extraction and $R = \rm FWHM_{\rm beam}/2 = 3.0$\,kpc as half the beam size, which we associate with the galaxy emission. We emphasize that this estimate carries significant uncertainties in both directions: the single Gaussian fit may underestimate the true line width (Figure~\ref{fig:line-spec}), while the beam-based size is a conservative upper limit on the true extent of \rubies, whose stellar effective radius is far more compact ($R_{\rm eff} = 209^{+34}_{-29}$\,pc; \citealt{weibel_2025}). We note that different derivations of \mdyn\ apply depending on the true distribution and geometry of the gas traced by \cii, currently unknown.\\

The range of \mgas\ estimates translates into corresponding intervals in gas fraction and depletion timescale $\tau_{\rm dep} = M_{\rm gas}/\rm SFR$ (Table~\ref{tab:physical_properties}). We find $f_{\rm mol} = M_{\rm mol}(\rm [CII],\,Z18) / M_\star \approx 20\%$, $f_{\rm HI} = M_{\rm HI}(\rm [CII],\,H21) / M_\star = 130$--$20\%$, and $f_{\rm gas} = M_{\rm gas}(\rm dust,\,P23) / M_\star < 163$--$9\%$, where each interval spans the low- to \highZ\ solutions. Using SFR estimates from SED modeling, we derive lower limits on the depletion timescale ranging from $\tau_{\rm dep} > 56\,\rm Myr$ (for $M_{\rm mol}(\rm [CII],\,Z18)$ and the highest $\rm SFR_{100}$) to $\tau_{\rm dep} > 0.7$--$1\,\rm Gyr$ (for the highest $M_{\rm HI}(\rm [CII],\,H21)$ and SFRs on 10 Myr timescales), the latter being comparable to or longer than the age of the universe at $z = 7.3$. With the exception of the highest $\rm SFR_{100}$ and lowest \mgas\ estimates, all solutions point to low star formation efficiencies and long depletion timescales.\\  
% Figure: gas fraction
\begin{figure*}
    \centering
    \includegraphics[width=\textwidth]{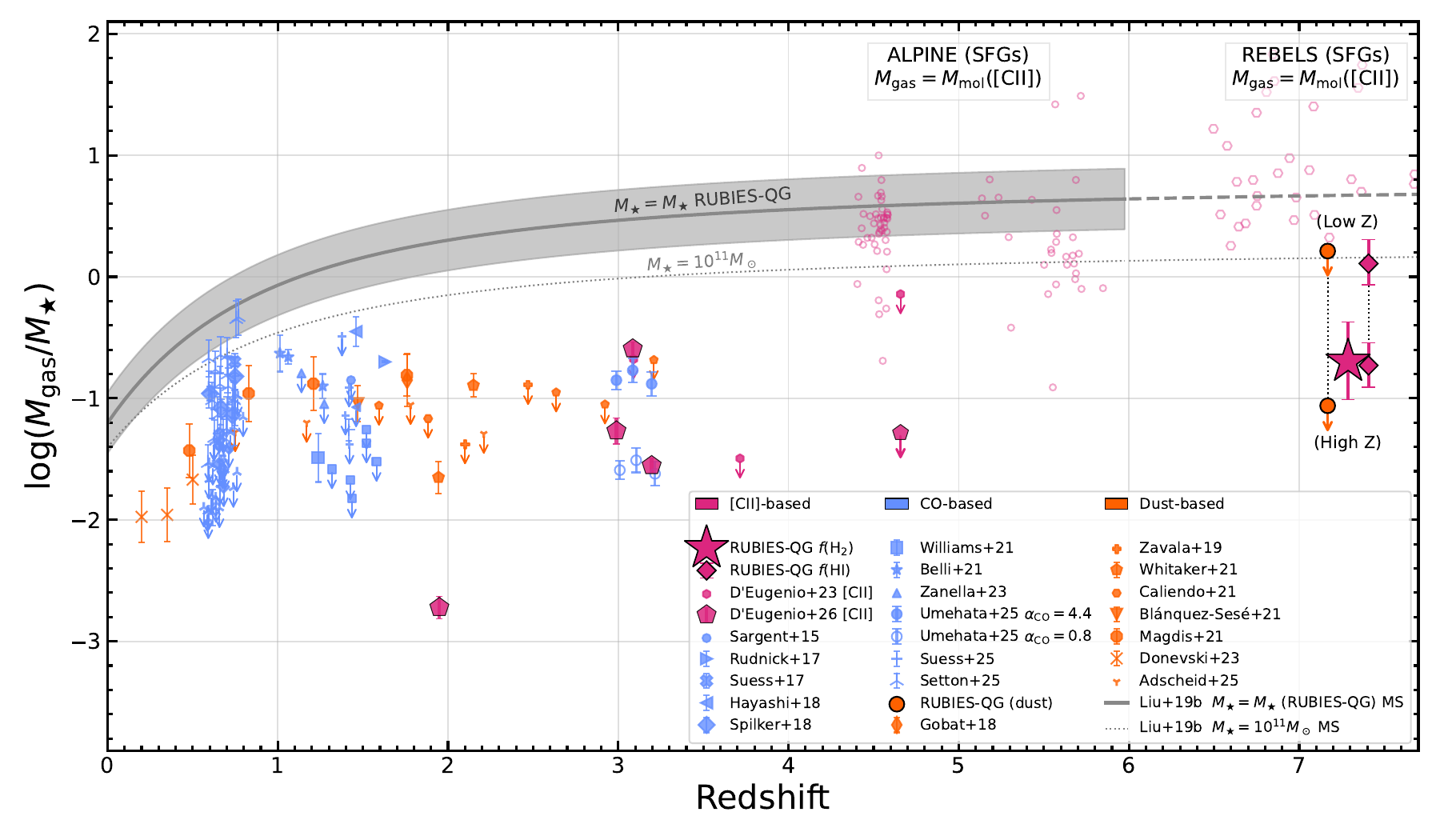}
    \caption{Gas fractions of quenched systems as a function of redshift. Purple, blue, and orange symbols indicate estimates based on \cii, CO, and dust, respectively; symbol shapes indicate the samples of QGs described in the legend. The gray solid line and shaded band show the locus of a main-sequence galaxy with stellar mass equal to that of \rubies\ ($M_\star = 10^{10.2}\,M_\odot$) parameterized as in \citet{liu_2019}; the dotted gray line shows the same trend for $M_\star = 10^{11}\,M_\odot$ for a better comparison with more massive low-redshift quiescent galaxies. For \rubies, the star marks $M_{\rm gas} = M_{\rm mol}({\rm [CII]})$ calibrated as in \cite{zanella_2018}; diamonds mark $M_{\rm gas} = M_{\rm HI}({\rm [CII]})$ following \cite{heintz_2021} for the two stellar metallicity solutions of \cite{weibel_2025}; orange circles mark upper limits on $M_{\rm gas}({\rm dust})$ for the same two solutions. Empty purple circles and hexagons indicate ALPINE and REBELS star-forming galaxies assuming $M_{\rm gas} = M_{\rm mol}({\rm [CII]})$ as in \cite{zanella_2018}. The compilation of molecular \fgas\ in the literature for quiescent galaxies is from \cite{deugenio_2026} and references therein.}
    \label{fig:fgas_z}
\end{figure*}

In Figure~\ref{fig:fgas_z}, we place these estimates in context by comparing \rubies\ with star-forming galaxies at similar redshift and quiescent galaxies across cosmic time. For a uniform comparison, we apply the $M_{\rm mol}(\rm [CII],\,Z18)$ calibration to \rubies, the ALPINE sources at $z\sim5$, and those from REBELS at $z \sim 7$, the same conversion as adopted in \cite{deugenio_2026} for their quiescent galaxies. As mentioned above, we consider only the \cii\ emission extracted in the central beam that we associate with the galaxy: including the gas emission on circumgalactic scales would increase the \fgas\ by a factor of $\approx3$. The molecular gas fraction of \rubies\ is $\approx 10\times$ lower than that of UV-bright star-forming galaxies at the same redshift, and similarly $\approx 10\times$ below the expected \fgas\ for a main-sequence galaxy with $M_\star = 10^{10.2}\,M_\odot$ extrapolated from \cite{liu_2019}. Therefore, while \rubies\ retains a substantial molecular gas reservoir after quenching, it is significantly more gas-poor than typical star-forming galaxies at $z \sim 7$. Comparing with quiescent galaxies at lower redshift, using the compilation of \cii-, CO-, and dust-based estimates from \cite{deugenio_2026}, the molecular gas fraction of \rubies\ is consistent with \cii-based results in that work and with CO- and dust-based estimates at lower redshift, with the latter declining only below $z \approx 1$. Taken together, these findings suggest that substantial cold gas reservoirs persist on galaxy scales and timescales of $\sim 100\,\rm Myr$ after quenching in early massive galaxies, pointing to at least a component of gas heating rather than complete removal as the mechanism responsible for suppressing star formation.

\subsection{Gas on circumgalactic scales and the role of outflows} 
\label{ssec:extended_halo}

So far, we have focused on the \cii\ emission from the
central beam, which we associate with the galaxy
itself. However, our analysis in
Section~\ref{ssec:mod-em} reveals \cii\ emission on
extended scales of $8\pm3$ kpc (Figure~\ref{fig:map_pv}),
much larger than the stellar continuum detected by JWST ($R_{\rm eff} = 209^{+33}_{-24}$\,pc;
\citealt{weibel_2025}). So-called ``\cii\ halos'' have
been reported around star-forming galaxies at high
redshift and attributed to several possible origins,
including faint satellites, neutral or ionized
circumgalactic gas, cold streams, outflows, or tidal
stripping \citep{fujimoto_2020_cii_halos,
rowland_2025}. \cite{deugenio_2026} also report
disturbed and possibly extended \cii\ and stellar
morphologies in their sample of quiescent galaxies at
lower redshift (see also \citealt{ji_2024} for details on GS-9209 at $z=4.6$), suggesting a role for mergers and tidal
interactions in extending and heating the residual gas.

Among these possible origins, an outflow scenario is
particularly compelling for \rubies. The extended \cii\
component is blueshifted with respect to the galaxy by
$\Delta\varv_{\rm [CII],\,ext} \approx -100$\,\kms\
(Figure~\ref{fig:map_pv}). This is consistent with the velocity 
offset reported by \cite{valentino_2025} for a tentative detection 
of blueshifted \mgii\ absorption in the rest-frame UV 
($\Delta\varv = -169 \pm 46$\,\kms\ with respect to the systemic 
redshift determined from JWST spectra), associated with a possible 
gas outflow. The \mgii\ velocity offset increases to $\approx230$ \kms\ with respect to the systemic redshift based on the \cii\ emission in the central beam (Table \ref{tab:line-flux}). \mgii\ broadly
traces the neutral phase of outflowing gas given its
ionization potential of 7.6\,eV, and under simple
geometric assumptions, the inferred energetics and mass
outflow rates, which appear extreme, are consistent with echoes of past AGN
activity, as found more broadly in recently quenched
galaxies at high redshift \citep{davies_2024, belli_2024, wu_2025,
zhu_2026_outflows, taylor_2026,
sun_2026_jades_outflows}. Applying the same
calibrations as in Section~\ref{ssec:gas_properties} to
the extended \cii\ component only, defined as the residual after 
subtracting the central galaxy emission from the total, we obtain $\log(M_{\rm
mol}^{\rm ext}/M_\odot) = 9.89^{+0.33}_{-0.32}$ and
$\log(M_{\rm HI}^{\rm ext}/M_\odot) = 10.70^{+0.21}_{-0.20}$
and $9.82^{+0.20}_{-0.20}$ for the low- and \highZ\
solutions, approximately 0.35 dex higher than the gas
mass associated with the galaxy alone. The gas mass
associated with the tentative \mgii\ outflow, $\log(M_{\rm
out}/M_\odot) \approx 9.1$ \citep{valentino_2025}, is of
a similar order of magnitude, further supporting a
physical connection and an origin of the extended \cii\
halo through episodes of gas expulsion from \rubies. It
is therefore tempting to associate the extended \cii\
emission with past AGN activity that expelled and
excited part of the gas. We note that the NIRSpec spectra show no clear
evidence for bright, AGN-powered emission lines, though
low-level activity cannot be excluded at the depth and
resolution of the available data. However, the mass and extreme properties of the possible outflows traced by \mgii\ and \cii\ are difficult to reconcile with any other origin.

Alternative scenarios appear less likely. The stellar
emission of \rubies\ is compact and symmetric,
almost point-like, with no clear residuals after
S\'{e}rsic modeling that would indicate a recent merger.
The available NIRCam images, among the deepest in the
literature, reveal no significant emission from putative
satellites in the immediate surroundings, though very
low surface brightness emitters, previously detected around massive quiescent galaxies at lower redshift \citep{suess_2023}, cannot
be entirely excluded at $z>7$. 

Regardless of its precise origin, given the low velocity offset of the extended emission, the mass of the system, and the depth of the potential well at this redshift, the cold gas is unlikely to escape the galaxy. In the absence of a mechanism able to maintain quiescence, a future reignition of star formation in \rubies\ appears plausible, rejuvenating the galaxy (as predicted by some recent models for distant quenched objects, \citealt{delucia_2024, fontanot_2025b, remus_2025_magneticum, lagos_2025, chandro-gomez_2025}), and bringing its properties closer to those of its possible massive descendants at lower redshift. And yet at the time of observation, \rubies\ remains quenched despite being immersed in a substantial reservoir of cold gas, implying that whatever mechanism is suppressing star formation must be remarkably effective at maintaining a low star formation efficiency on $\sim100$\,Myr timescales even in the presence of abundant fuel.

\subsection{Testable scenarios and assumptions}
\label{ssec:caveats-tests}

Several key assumptions underlie the physical scenarios discussed above. We address the most important ones here and outline
how they can be tested with forthcoming observations.

The first caveat concerns the spatial decomposition
of the \cii\ emission.
Throughout this work, we assumed that the emission
from the central beam is associated with the galaxy,
which is a conservative upper limit given that some
flux from the extended component is likely included.
Conversely, the true flux and size of the extended
halo remain limited by the current S/N.
Both issues can be addressed with deeper observations
along two complementary lines: high-resolution
interferometric imaging, matched to the resolution
of the NIRCam stellar emission, will separate the
galaxy from its surroundings; deeper low-resolution
observations will push the detectability of the
extended halo.
Combining the two --- recovering the total flux while
resolving it in multiple independent beams --- will
also enable a first study of the gas kinematics and geometry of the system,
currently limited to the blue- and redshifted
decomposition in Figure~\ref{fig:map_pv}, and place
independent constraints on the origin of the emission
across interstellar and circumgalactic scales,
including rotation, inflows, outflows, and merger
signatures.

A second set of caveats concerns the conversion from
\lcii\ to gas mass.
The assumption that \cii\ primarily traces the
neutral and molecular phases can be challenged if the
ionized phase contributes substantially.
Critically, the relative contributions of the three
phases likely differ between the galaxy and
circumgalactic scales: while a conversion to
molecular gas is physically motivated for the
galaxy-scale emission,
a neutral gas calibration, with a possible significant contribution from the ionized component, may be more appropriate
for the extended halo, where no discernible trace
of recent star formation is found.
The metallicity dependence of the calibrations adds
a further source of uncertainty in the absence of a
direct gas-phase metallicity measurement.

Forthcoming deep JWST/NIRSpec spectra with medium-resolution gratings (GO-7488, PI: A.~Weibel;
GO-12396, PIs: R.~Naidu \& A.~de Graaff) hold the
promise of addressing several of these limitations at
once.
The improved S/N and spectral resolution will allow
the detection of rest-frame optical emission lines
from the ionized phase, constraining their origin
(residual star formation, shocks, or AGN activity)
and providing a direct measurement of the gas-phase
metallicity.
Combined modeling of rest-frame UV and optical
emission lines with photoionization codes, together
with additional fine-structure lines accessible
with ALMA for this source (e.g.,
[\ion{N}{II}] 122 and 205\,$\mu$m), will constrain
the contribution from the ionized gas phase.
Detection of ionized gas on extended scales
(e.g., \oiii\,$\lambda5007$), not unusual around
AGN \citep[e.g.,][for a massive quiescent galaxy at $z=3.7$]{perez-gonzalez_2025}, would
further support a significant ionized contribution
to the \cii\ emission.
Follow-up of dense neutral gas tracers with ALMA
(e.g., [\ion{O}{I}] 145\,$\mu$m) will provide an
independent handle on the neutral phase.
Deeper spectra around \mgii\ will independently test the tentative
outflow detection of \cite{valentino_2025}; combined with higher-resolution ALMA observations constraining the co-spatiality and
geometry of the \mgii\ absorption and extended \cii\ emission, this
will help corroborate or rule out past AGN activity as the origin
of the extended \cii\ emission.

Finally, multi-wavelength follow-up of the dust
continuum could help lift the assumptions on dust
temperature and emissivity index adopted in
Section~\ref{ssec:dust}. A better-constrained dust mass, coupled with gas mass values, will provide an independent estimate of the gas-to-dust ratio -- a quantity that may differ substantially between
star-forming galaxies and quenched or post-starburst
systems \citep{whitaker_2021, lorenzon_2025, spilker_2025}. 
Additionally, high-frequency observations of the dust continuum could help constrain the SFH of the galaxy, as SED models produce slightly different predictions for \lir\ (Section \ref{ssec:dust}).

\section{Conclusions}
\label{sec:conclusions}
We report the detection of \cii\ 158\,$\mu$m emission
in and around \rubies\
at $z = 7.27$, the most distant massive quenched
galaxy known to date \citep{weibel_2025}.
The emission extends over $8 \pm 3$\,kpc, well
beyond the compact stellar body, with a
significant fraction of $\approx70\%$ arising from the gaseous halo on circumgalactic scales.
No dust continuum is detected at 1.3\,mm.
Our main findings are as follows.

\begin{itemize}

\item Cold gas reservoir after quenching:
Despite having suppressed star formation for $\sim 100$\,Myr, \rubies\ retains a substantial
cold gas reservoir.
Depending on the assumed calibration and stellar
metallicity solution, we find
$\log(M_{\rm mol}/M_\odot) = 9.53^{+0.32}_{-0.31}$ and
$\log(M_{\rm HI}/M_\odot) = 9.46$--$10.34$,
corresponding to gas fractions
of $f_{\rm gas} \approx20\%$ (molecular) or $\approx 20$--$130\%$ (neutral).
Using the \cii-to-molecular gas conversion as a reference,
the molecular gas fraction of \rubies\ is $\approx 10\times$
lower than those of UV-bright star-forming galaxies
at $z \sim 7$ and of main-sequence galaxies of
similar mass extrapolated from well-established
scaling relations, yet consistent with the gas
fractions of quiescent galaxies across a wide
redshift range down to $z \approx 1$.

\item Low star formation efficiency:
The \lcii\ luminosity of \rubies\ is broadly
consistent with the \lcii--SFR relation when
considering SFR estimates from SED modeling averaged over 100\,Myr,
but inconsistent with the much lower SFRs inferred
on 10\,Myr timescales.
Depletion timescales range from
$\tau_{\rm dep} > 56$\,Myr to
$\tau_{\rm dep} > 0.7$--$1$\,Gyr, the latter comparable to
or exceeding the age of the universe at $z = 7.3$.
Based on the shortest-timescale SFR estimates, the available gas reservoirs imply a low star formation efficiency in the system.

\item Gas on circumgalactic scales:
The \cii\ emission extends significantly beyond the
stellar body of \rubies, with the extended component
carrying twice as much gas mass as the
galaxy alone under the same assumptions on the \lcii-to-\mgas\ conversion factor. The stellar morphology shows no evidence of recent mergers or satellites at the current depth, and the existing NIRSpec spectra, albeit limited in S/N and resolution, show no clear signs of ongoing strong nuclear activity. Yet, the velocity offset of the extended \cii\ component is
consistent with the tentative \mgii\ absorption
outflow reported by \cite{valentino_2025}, suggesting
that past AGN activity may have expelled and excited
the circumgalactic gas.

\item Quenching at cosmic dawn:
The combination of a compact stellar structure,
suppressed star formation, and a large, possibly
AGN-ejected and heated gas reservoir at $z = 7.3$ --- just
700\,Myr after the Big Bang --- implies that the
mechanisms halting star formation act without fully
expelling or depleting the cold gas from the system, and are already in place
at the earliest cosmic epochs probed to date. 

\end{itemize}

Taken together, these results indicate that \rubies\ is immersed in a sizeable reservoir of cold gas despite showing no sign of ongoing star formation, a 
puzzle that future observations will help to unravel,
with implications for our understanding of galaxy
quenching in the early Universe.

\begin{acknowledgements}
We warmly thank Chiara D'Eugenio for kindly providing the literature compilation shown in Figure \ref{fig:fgas_z} and information about her sample, and Mariko Kubo for sharing the photometry of QG1--3. FV, KI, AP, and PZ acknowledge support from the Independent Research Fund Denmark (DFF) under grant 3120-00043B. AdG acknowledges support from a Clay Fellowship awarded by the Smithsonian Astrophysical Observatory. This work has received funding from the Swiss State Secretariat for Education, Research and Innovation (SERI) under contract number MB22.00072, as well as from the Swiss National Science Foundation (SNSF) through project grants 200020\_207349 and 2000-1-243073. This work is based in part on observations made with the NASA/ESA/CSA James Webb Space Telescope. The data were obtained from the Mikulski Archive for Space Telescopes at the Space Telescope Science Institute, which is operated by the Association of Universities for Research in Astronomy, Inc., under NASA contract NAS 5-03127 for JWST. Some of the data products presented herein were retrieved from the Dawn JWST Archive (DJA). DJA is an initiative of the Cosmic Dawn Center, which is funded by the Danish National Research Foundation under grant DNRF140. 

\end{acknowledgements}

\bibliographystyle{aa} 
\bibliography{bib_cii, MyBib_over_latest_AP}

\begin{thebibliography}{130}
\expandafter\ifx\csname natexlab\endcsname\relax\def\natexlab#1{#1}\fi

\bibitem[{{Akhshik} {et~al.}(2023){Akhshik}, {Whitaker}, {Leja}, {Richard}, {Spilker}, {Song}, {Brammer}, {Bezanson}, {Ebeling}, {Gallazzi}, {Mahler}, {Mowla}, {Nelson}, {Pacifici}, {Sharon}, {Toft}, {Williams}, {Wright}, \& {Zabl}}]{akhshik_2023}
{Akhshik}, M., {Whitaker}, K.~E., {Leja}, J., {et~al.} 2023, \apj, 943, 179

\bibitem[{{Alberts} {et~al.}(2024){Alberts}, {Williams}, {Helton}, {Suess}, {Ji}, {Shivaei}, {Lyu}, {Rieke}, {Baker}, {Bonaventura}, {Bunker}, {Carniani}, {Charlot}, {Curtis-Lake}, {D'Eugenio}, {Eisenstein}, {de Graaff}, {Hainline}, {Hausen}, {Johnson}, {Maiolino}, {Parlanti}, {Rieke}, {Robertson}, {Sun}, {Tacchella}, {Willmer}, \& {Willott}}]{alberts_2024}
{Alberts}, S., {Williams}, C.~C., {Helton}, J.~M., {et~al.} 2024, \apj, 975, 85

\bibitem[{{Baker} {et~al.}(2025{\natexlab{a}}){Baker}, {Lim}, {D'Eugenio}, {Maiolino}, {Ji}, {Arribas}, {Bunker}, {Carniani}, {Charlot}, {de Graaff}, {Hainline}, {Looser}, {Lyu}, {Rinaldi}, {Robertson}, {Schaller}, {Schaye}, {Scholtz}, {{\"U}bler}, {Williams}, {Willmer}, {Willott}, \& {Zhu}}]{baker_2025}
{Baker}, W.~M., {Lim}, S., {D'Eugenio}, F., {et~al.} 2025{\natexlab{a}}, \mnras, 539, 557

\bibitem[{{Baker} {et~al.}(2025{\natexlab{b}}){Baker}, {Valentino}, {Lagos}, {Ito}, {Jespersen}, {Gottumukkala}, {Hjorth}, {Langeroodi}, \& {Sedgewick}}]{baker_2025_dja}
{Baker}, W.~M., {Valentino}, F., {Lagos}, C. d.~P., {et~al.} 2025{\natexlab{b}}, \aap, 702, A270

\bibitem[{{Baron} {et~al.}(2023){Baron}, {Netzer}, {French}, {Lutz}, {Davies}, \& {Prochaska}}]{baron2023}
{Baron}, D., {Netzer}, H., {French}, K.~D., {et~al.} 2023, \mnras, 524, 2741

\bibitem[{{Belli} {et~al.}(2021){Belli}, {Contursi}, {Genzel}, {Tacconi}, {F{\"o}rster-Schreiber}, {Lutz}, {Combes}, {Neri}, {Garc{\'\i}a-Burillo}, {Schuster}, {Herrera-Camus}, {Tadaki}, {Davies}, {Davies}, {Johnson}, {Lee}, {Leja}, {Nelson}, {Price}, {Shangguan}, {Shimizu}, {Tacchella}, \& {{\"U}bler}}]{belli_2021}
{Belli}, S., {Contursi}, A., {Genzel}, R., {et~al.} 2021, \apjl, 909, L11

\bibitem[{{Belli} {et~al.}(2024){Belli}, {Park}, {Davies}, {Mendel}, {Johnson}, {Conroy}, {Benton}, {Bugiani}, {Emami}, {Leja}, {Li}, {Maheson}, {Mathews}, {Naidu}, {Nelson}, {Tacchella}, {Terrazas}, \& {Weinberger}}]{belli_2024}
{Belli}, S., {Park}, M., {Davies}, R.~L., {et~al.} 2024, \nat, 630, 54

\bibitem[{{B{\'e}thermin} {et~al.}(2020){B{\'e}thermin}, {Fudamoto}, {Ginolfi}, {Loiacono}, {Khusanova}, {Capak}, {Cassata}, {Faisst}, {Le F{\`e}vre}, {Schaerer}, {Silverman}, {Yan}, {Amorin}, {Bardelli}, {Boquien}, {Cimatti}, {Davidzon}, {Dessauges-Zavadsky}, {Fujimoto}, {Gruppioni}, {Hathi}, {Ibar}, {Jones}, {Koekemoer}, {Lagache}, {Lemaux}, {Moreau}, {Oesch}, {Pozzi}, {Riechers}, {Talia}, {Toft}, {Vallini}, {Vergani}, {Zamorani}, \& {Zucca}}]{Bethermin+2020}
{B{\'e}thermin}, M., {Fudamoto}, Y., {Ginolfi}, M., {et~al.} 2020, \aap, 643, A2

\bibitem[{{Bezanson} {et~al.}(2022){Bezanson}, {Labbe}, {Whitaker}, {Leja}, {Price}, {Franx}, {Brammer}, {Marchesini}, {Zitrin}, {Wang}, {Weaver}, {Furtak}, {Atek}, {Coe}, {Cutler}, {Dayal}, {van Dokkum}, {Feldmann}, {Forster Schreiber}, {Fujimoto}, {Geha}, {Glazebrook}, {de Graaff}, {Greene}, {Juneau}, {Kassin}, {Kriek}, {Khullar}, {Maseda}, {Mowla}, {Muzzin}, {Nanayakkara}, {Nelson}, {Oesch}, {Pacifici}, {Pan}, {Papovich}, {Setton}, {Shapley}, {Smit}, {Stefanon}, {Taylor}, \& {Williams}}]{bezanson_2022}
{Bezanson}, R., {Labbe}, I., {Whitaker}, K.~E., {et~al.} 2022, arXiv e-prints, arXiv:2212.04026

\bibitem[{{Bl{\'a}nquez-Ses{\'e}} {et~al.}(2023){Bl{\'a}nquez-Ses{\'e}}, {G{\'o}mez-Guijarro}, {Magdis}, {Magnelli}, {Gobat}, {Daddi}, {Franco}, {Whitaker}, {Valentino}, {Adscheid}, {Schinnerer}, {Zanella}, {Xiao}, {Wang}, {Liu}, {Kokorev}, \& {Elbaz}}]{blanquez-sese_2023}
{Bl{\'a}nquez-Ses{\'e}}, D., {G{\'o}mez-Guijarro}, C., {Magdis}, G.~E., {et~al.} 2023, \aap, 674, A166

\bibitem[{{Bouwens} {et~al.}(2021){Bouwens}, {Smit}, {Schouws}, {Stefanon}, {Bowler}, {Endsley}, {Gonzalez}, {Inami}, {Stark}, {Oesch}, {Hodge}, {Aravena}, {da Cunha}, {Dayal}, {de Looze}, {Ferrara}, {Fudamoto}, {Graziani}, {Li}, {Nanayakkara}, {Pallotini}, {Schneider}, {Sommovigo}, {Topping}, {van der Werf}, {Barrufet}, {Hygate}, {Labbe}, {Riechers}, \& {Witstok}}]{bouwens_2021}
{Bouwens}, R.~J., {Smit}, R., {Schouws}, S., {et~al.} 2021, arXiv e-prints, arXiv:2106.13719

\bibitem[{{Carnall} {et~al.}(2022){Carnall}, {McLeod}, {McLure}, {Dunlop}, {Begley}, {Cullen}, {Donnan}, {Hamadouche}, {Jewell}, {Jones}, {Pollock}, \& {Wild}}]{carnall_2022}
{Carnall}, A.~C., {McLeod}, D.~J., {McLure}, R.~J., {et~al.} 2022, arXiv e-prints, arXiv:2208.00986

\bibitem[{{Carnall} {et~al.}(2023){Carnall}, {McLure}, {Dunlop}, {McLeod}, {Wild}, {Cullen}, {Magee}, {Begley}, {Cimatti}, {Donnan}, {Hamadouche}, {Jewell}, \& {Walker}}]{carnall_2023b}
{Carnall}, A.~C., {McLure}, R.~J., {Dunlop}, J.~S., {et~al.} 2023, \nat, 619, 716

\bibitem[{{Chabrier}(2003)}]{chabrier_2003}
{Chabrier}, G. 2003, \pasp, 115, 763

\bibitem[{{Chaikin} {et~al.}(2026){Chaikin}, {Schaye}, {Hu{\v{s}}ko}, {Lacey}, {Ploeckinger}, \& {Schaller}}]{chaikin_2026}
{Chaikin}, E., {Schaye}, J., {Hu{\v{s}}ko}, F., {et~al.} 2026, arXiv e-prints, arXiv:2601.15207

\bibitem[{{Chandro-G{\'o}mez} {et~al.}(2025){Chandro-G{\'o}mez}, {Lagos}, {Power}, {Baker}, {Ben{\'\i}tez-Llambay}, {Chaikin}, {Chittenden}, {Correa}, {Frenk}, {Hu{\v{s}}ko}, {McGibbon}, {Nanayakkara}, {Ploeckinger}, {Richings}, {Schaller}, {Schaye}, \& {Trayford}}]{chandro-gomez_2025}
{Chandro-G{\'o}mez}, {\'A}., {Lagos}, C. d.~P., {Power}, C., {et~al.} 2025, arXiv e-prints, arXiv:2512.16208

\bibitem[{{Choi} {et~al.}(2016){Choi}, {Dotter}, {Conroy}, {Cantiello}, {Paxton}, \& {Johnson}}]{choi_2016}
{Choi}, J., {Dotter}, A., {Conroy}, C., {et~al.} 2016, \apj, 823, 102

\bibitem[{{Conroy} \& {Gunn}(2010)}]{conroy_2010}
{Conroy}, C. \& {Gunn}, J.~E. 2010, \apj, 712, 833

\bibitem[{{Cornwell}(2008)}]{Cornwell2008}
{Cornwell}, T.~J. 2008, IEEE Journal of Selected Topics in Signal Processing, 2, 793

\bibitem[{{da Cunha} {et~al.}(2013){da Cunha}, {Groves}, {Walter}, {Decarli}, {Weiss}, {Bertoldi}, {Carilli}, {Daddi}, {Elbaz}, {Ivison}, {Maiolino}, {Riechers}, {Rix}, {Sargent}, \& {Smail}}]{dacunha_2013}
{da Cunha}, E., {Groves}, B., {Walter}, F., {et~al.} 2013, \apj, 766, 13

\bibitem[{{Davies} {et~al.}(2024){Davies}, {Belli}, {Park}, {Mendel}, {Johnson}, {Conroy}, {Benton}, {Bugiani}, {Emami}, {Leja}, {Li}, {Maheson}, {Mathews}, {Naidu}, {Nelson}, {Tacchella}, {Terrazas}, \& {Weinberger}}]{davies_2024}
{Davies}, R.~L., {Belli}, S., {Park}, M., {et~al.} 2024, \mnras, 528, 4976

\bibitem[{{Davis} {et~al.}(2011){Davis}, {Alatalo}, {Sarzi}, {Bureau}, {Young}, {Blitz}, {Serra}, {Crocker}, {Krajnovi{\'c}}, {McDermid}, {Bois}, {Bournaud}, {Cappellari}, {Davies}, {Duc}, {de Zeeuw}, {Emsellem}, {Khochfar}, {Kuntschner}, {Lablanche}, {Morganti}, {Naab}, {Oosterloo}, {Scott}, \& {Weijmans}}]{davis_2011}
{Davis}, T.~A., {Alatalo}, K., {Sarzi}, M., {et~al.} 2011, \mnras, 417, 882

\bibitem[{{de Graaff} {et~al.}(2025{\natexlab{a}}){de Graaff}, {Brammer}, {Weibel}, {Lewis}, {Maseda}, {Oesch}, {Bezanson}, {Boogaard}, {Cleri}, {Cooper}, {Gottumukkala}, {Greene}, {Hirschmann}, {Hviding}, {Katz}, {Labb{\'e}}, {Leja}, {Matthee}, {McConachie}, {Miller}, {Naidu}, {Price}, {Rix}, {Setton}, {Suess}, {Wang}, {Whitaker}, \& {Williams}}]{degraaff_2024_rubies}
{de Graaff}, A., {Brammer}, G., {Weibel}, A., {et~al.} 2025{\natexlab{a}}, \aap, 697, A189

\bibitem[{{de Graaff} {et~al.}(2025{\natexlab{b}}){de Graaff}, {Setton}, {Brammer}, {Cutler}, {Suess}, {Labb{\'e}}, {Leja}, {Weibel}, {Maseda}, {Whitaker}, {Bezanson}, {Boogaard}, {Cleri}, {De Lucia}, {Franx}, {Greene}, {Hirschmann}, {Matthee}, {McConachie}, {Naidu}, {Oesch}, {Price}, {Rix}, {Valentino}, {Wang}, \& {Williams}}]{deGraaff_2025}
{de Graaff}, A., {Setton}, D.~J., {Brammer}, G., {et~al.} 2025{\natexlab{b}}, Nature Astronomy, 9, 280

\bibitem[{{De Looze} {et~al.}(2014){De Looze}, {Cormier}, {Lebouteiller}, {Madden}, {Baes}, {Bendo}, {Boquien}, {Boselli}, {Clements}, {Cortese}, {Cooray}, {Galametz}, {Galliano}, {Graci{\'a}-Carpio}, {Isaak}, {Karczewski}, {Parkin}, {Pellegrini}, {R{\'e}my-Ruyer}, {Spinoglio}, {Smith}, \& {Sturm}}]{delooze_2014}
{De Looze}, I., {Cormier}, D., {Lebouteiller}, V., {et~al.} 2014, \aap, 568, A62

\bibitem[{{De Lucia} {et~al.}(2025){De Lucia}, {Fontanot}, {Hirschmann}, \& {Xie}}]{delucia_2025_book_quenching}
{De Lucia}, G., {Fontanot}, F., {Hirschmann}, M., \& {Xie}, L. 2025, arXiv e-prints, arXiv:2502.01724

\bibitem[{{De Lucia} {et~al.}(2024){De Lucia}, {Fontanot}, {Xie}, \& {Hirschmann}}]{delucia_2024}
{De Lucia}, G., {Fontanot}, F., {Xie}, L., \& {Hirschmann}, M. 2024, \aap, 687, A68

\bibitem[{{De Vis} {et~al.}(2019){De Vis}, {Jones}, {Viaene}, {Casasola}, {Clark}, {Baes}, {Bianchi}, {Cassara}, {Davies}, {De Looze}, {Galametz}, {Galliano}, {Lianou}, {Madden}, {Manilla-Robles}, {Mosenkov}, {Nersesian}, {Roychowdhury}, {Xilouris}, \& {Ysard}}]{de-vis_2019}
{De Vis}, P., {Jones}, A., {Viaene}, S., {et~al.} 2019, \aap, 623, A5

\bibitem[{{D'Eugenio} {et~al.}(2026){D'Eugenio}, {Daddi}, {Gobat}, {Jin}, {Liu}, {Sun}, {Gentile}, {Bruckmann}, {Liu}, {Delvecchio}, {Vallini}, {Magnelli}, \& {Zanella}}]{deugenio_2026}
{D'Eugenio}, C., {Daddi}, E., {Gobat}, R., {et~al.} 2026, arXiv e-prints, arXiv:2604.09347

\bibitem[{{D'Eugenio} {et~al.}(2023){D'Eugenio}, {Daddi}, {Liu}, \& {Gobat}}]{deugenio_2023}
{D'Eugenio}, C., {Daddi}, E., {Liu}, D., \& {Gobat}, R. 2023, \aap, 678, L9

\bibitem[{{Donevski} {et~al.}(2023){Donevski}, {Damjanov}, {Nanni}, {Man}, {Giulietti}, {Romano}, {Lapi}, {Narayanan}, {Dav{\'e}}, {Shivaei}, {Sohn}, {Junais}, {Pantoni}, \& {Li}}]{Donevski+2023}
{Donevski}, D., {Damjanov}, I., {Nanni}, A., {et~al.} 2023, \aap, 678, A35

\bibitem[{{Donnan} {et~al.}(2024){Donnan}, {McLure}, {Dunlop}, {McLeod}, {Magee}, {Arellano-C{\'o}rdova}, {Barrufet}, {Begley}, {Bowler}, {Carnall}, {Cullen}, {Ellis}, {Fontana}, {Illingworth}, {Grogin}, {Hamadouche}, {Koekemoer}, {Liu}, {Mason}, {Santini}, \& {Stanton}}]{donnan_2024}
{Donnan}, C.~T., {McLure}, R.~J., {Dunlop}, J.~S., {et~al.} 2024, \mnras, 533, 3222

\bibitem[{{D'Onofrio} {et~al.}(2026){D'Onofrio}, {Spilker}, {Bezanson}, {Feldmann}, {Goulding}, {Greene}, {Kriek}, {Kumar}, {Luo}, {Narayanan}, {Setton}, {Suess}, \& {Verrico}}]{donofrio_2026}
{D'Onofrio}, V.~R., {Spilker}, J.~S., {Bezanson}, R., {et~al.} 2026, \apj, 1003, 52

\bibitem[{{Ellison} {et~al.}(2025){Ellison}, {Huang}, {Yang}, {Wang}, {Wild}, {Rasmussen}, {Jimenez-Donaire}, {Rowlands}, {Wilkinson}, {Brown}, \& {Leung}}]{ellison_2025_hi_psb}
{Ellison}, S., {Huang}, Q., {Yang}, D., {et~al.} 2025, The Open Journal of Astrophysics, 8, 87

\bibitem[{{Faisst} {et~al.}(2020){Faisst}, {Fudamoto}, {Oesch}, {Scoville}, {Riechers}, {Pavesi}, \& {Capak}}]{Faisst+2020}
{Faisst}, A.~L., {Fudamoto}, Y., {Oesch}, P.~A., {et~al.} 2020, \mnras, 498, 4192

\bibitem[{{Farcy} {et~al.}(2025){Farcy}, {Hirschmann}, {Somerville}, {Choi}, {Koudmani}, {Naab}, {Weinberger}, {Bennett}, {Bhowmick}, {Choi}, {Hernquist}, {Hlavacek-Larrondo}, {Terrazas}, \& {Valentino}}]{farcy_2025}
{Farcy}, M., {Hirschmann}, M., {Somerville}, R.~S., {et~al.} 2025, \mnras\ (submitted), arXiv:2504.08041

\bibitem[{{Fontanot} {et~al.}(2025){Fontanot}, {Decarli}, {De Lucia}, {Cucciati}, {Xie}, \& {Hirschmann}}]{fontanot_2025b}
{Fontanot}, F., {Decarli}, R., {De Lucia}, G., {et~al.} 2025, arXiv e-prints, arXiv:2511.15789

\bibitem[{{Foreman-Mackey} {et~al.}(2013){Foreman-Mackey}, {Hogg}, {Lang}, \& {Goodman}}]{Foreman+2013}
{Foreman-Mackey}, D., {Hogg}, D.~W., {Lang}, D., \& {Goodman}, J. 2013, Publications of the Astronomical Society of the Pacific, 125, 306

\bibitem[{{French} {et~al.}(2023){French}, {Smercina}, {Rowlands}, {Tripathi}, {Zabludoff}, {Smith}, {Narayanan}, {Yang}, {Shirley}, \& {Alatalo}}]{french2023}
{French}, K.~D., {Smercina}, A., {Rowlands}, K., {et~al.} 2023, \apj, 942, 25

\bibitem[{{French} {et~al.}(2015){French}, {Yang}, {Zabludoff}, {Narayanan}, {Shirley}, {Walter}, {Smith}, \& {Tremonti}}]{french_2015}
{French}, K.~D., {Yang}, Y., {Zabludoff}, A., {et~al.} 2015, \apj, 801, 1

\bibitem[{{French} {et~al.}(2018){French}, {Zabludoff}, {Yoon}, {Shirley}, {Yang}, {Smercina}, {Smith}, \& {Narayanan}}]{french_2018}
{French}, K.~D., {Zabludoff}, A.~I., {Yoon}, I., {et~al.} 2018, \apj, 861, 123

\bibitem[{{Fudamoto} {et~al.}(2025){Fudamoto}, {Inoue}, {Bouwens}, {Inami}, {Smit}, {Stark}, {Aravena}, {Pallottini}, {Hashimoto}, {Oguri}, {Bowler}, {da Cunha}, {Dayal}, {Ferrara}, {Fujimoto}, {Heintz}, {Hygate}, {van Leeuwen}, {De Looze}, {Rowland}, {Stefanon}, {Sugahara}, {Witstok}, \& {van der Werf}}]{fudamoto_2025}
{Fudamoto}, Y., {Inoue}, A.~K., {Bouwens}, R., {et~al.} 2025, arXiv e-prints, arXiv:2504.03831

\bibitem[{{Fujimoto} {et~al.}(2020){Fujimoto}, {Silverman}, {Bethermin}, {Ginolfi}, {Jones}, {Le F{\`e}vre}, {Dessauges-Zavadsky}, {Rujopakarn}, {Faisst}, {Fudamoto}, {Cassata}, {Morselli}, {Maiolino}, {Schaerer}, {Capak}, {Yan}, {Vallini}, {Toft}, {Loiacono}, {Zamorani}, {Talia}, {Narayanan}, {Hathi}, {Lemaux}, {Boquien}, {Amorin}, {Ibar}, {Koekemoer}, {M{\'e}ndez-Hern{\'a}ndez}, {Bardelli}, {Vergani}, {Zucca}, {Romano}, \& {Cimatti}}]{fujimoto_2020_cii_halos}
{Fujimoto}, S., {Silverman}, J.~D., {Bethermin}, M., {et~al.} 2020, \apj, 900, 1

\bibitem[{{Gobat} {et~al.}(2018){Gobat}, {Daddi}, {Magdis}, {Bournaud}, {Sargent}, {Martig}, {Jin}, {Finoguenov}, {B{\'e}thermin}, {Hwang}, {Renzini}, {Wilson}, {Aretxaga}, {Yun}, {Strazzullo}, \& {Valentino}}]{gobat_2018}
{Gobat}, R., {Daddi}, E., {Magdis}, G., {et~al.} 2018, Nature Astronomy, 2, 239

\bibitem[{{Heintz} {et~al.}(2025{\natexlab{a}}){Heintz}, {Brammer}, {Watson}, {Oesch}, {Keating}, {Hayes}, {Abdurro'uf}, {Arellano-C{\'o}rdova}, {Carnall}, {Christiansen}, {Cullen}, {Dav{\'e}}, {Dayal}, {Ferrara}, {Finlator}, {Fynbo}, {Flury}, {Gelli}, {Gillman}, {Gottumukkala}, {Gould}, {Greve}, {Hardin}, {Hsiao}, {Hutter}, {Jakobsson}, {Killi}, {Khosravaninezhad}, {Laursen}, {Lee}, {Magdis}, {Matthee}, {Naidu}, {Narayanan}, {Pollock}, {Prescott}, {Rusakov}, {Shuntov}, {Sneppen}, {Smit}, {Tanvir}, {Terp}, {Toft}, {Valentino}, {Vijayan}, {Weaver}, {Wise}, \& {Witstok}}]{heintz_2025}
{Heintz}, K.~E., {Brammer}, G.~B., {Watson}, D., {et~al.} 2025{\natexlab{a}}, \aap, 693, A60

\bibitem[{{Heintz} {et~al.}(2023){Heintz}, {De Cia}, {Th{\"o}ne}, {Krogager}, {Yates}, {Vejlgaard}, {Konstantopoulou}, {Fynbo}, {Watson}, {Narayanan}, {Wilson}, {Arabsalmani}, {Campana}, {D'Elia}, {De Pasquale}, {Hartmann}, {Izzo}, {Jakobsson}, {Kouveliotou}, {Levan}, {Li}, {Malesani}, {Melandri}, {Milvang-Jensen}, {M{\o}ller}, {Palazzi}, {Palmerio}, {Petitjean}, {Pugliese}, {Rossi}, {Saccardi}, {Salvaterra}, {Savaglio}, {Schady}, {Stratta}, {Tanvir}, {de Ugarte Postigo}, {Vergani}, {Wiersema}, {Wijers}, \& {Zafar}}]{heintz_2023_dtg}
{Heintz}, K.~E., {De Cia}, A., {Th{\"o}ne}, C.~C., {et~al.} 2023, \aap, 679, A91

\bibitem[{{Heintz} {et~al.}(2021){Heintz}, {Watson}, {Oesch}, {Narayanan}, \& {Madden}}]{heintz_2021}
{Heintz}, K.~E., {Watson}, D., {Oesch}, P.~A., {Narayanan}, D., \& {Madden}, S.~C. 2021, \apj, 922, 147

\bibitem[{{Heintz} {et~al.}(2025{\natexlab{b}}){Heintz}, {Watson}, {Valentino}, {Gottumukkala}, {Narayanan}, {Yates}, {Terp}, {Nezhad}, {Weaver}, {Witstok}, {Brammer}, {Andersen}, {Sneppen}, {Pollock}, {Algera}, {Rowland}, {Oesch}, {Magdis}, {Nikopoulos}, \& {Knudsen}}]{heintz_2025_inefficient_dust_production}
{Heintz}, K.~E., {Watson}, D., {Valentino}, F., {et~al.} 2025{\natexlab{b}}, arXiv e-prints, arXiv:2510.07936

\bibitem[{{Herrera-Camus} {et~al.}(2025){Herrera-Camus}, {Gonz{\'a}lez-L{\'o}pez}, {F{\"o}rster Schreiber}, {Aravena}, {de Looze}, {Spilker}, {Tadaki}, {Barcos-Mu{\~n}oz}, {Assef}, {Birkin}, {Bolatto}, {Bouwens}, {Bovino}, {Bowler}, {Calistro Rivera}, {da Cunha}, {Davies}, {Davies}, {D{\'\i}az-Santos}, {Ferrara}, {Fisher}, {Genzel}, {Hodge}, {Ikeda}, {Killi}, {Lee}, {Li}, {Li}, {Liu}, {Lutz}, {Mitsuhashi}, {Narayanan}, {Naab}, {Palla}, {Price}, {Posses}, {Rela{\~n}o}, {Smit}, {Solimano}, {Sternberg}, {Tacconi}, {Telikova}, {{\"U}bler}, {van der Giessen}, {Veilleux}, {Villanueva}, \& {Baeza-Garay}}]{herrera-camus_2025}
{Herrera-Camus}, R., {Gonz{\'a}lez-L{\'o}pez}, J., {F{\"o}rster Schreiber}, N., {et~al.} 2025, \aap, 699, A80

\bibitem[{{Hirashita} {et~al.}(2014){Hirashita}, {Ferrara}, {Dayal}, \& {Ouchi}}]{hirashita_2014}
{Hirashita}, H., {Ferrara}, A., {Dayal}, P., \& {Ouchi}, M. 2014, \mnras, 443, 1704

\bibitem[{{Hunter} {et~al.}(2023){Hunter}, {Indebetouw}, {Brogan}, {Berry}, {Chang}, {Francke}, {Geers}, {G{\'o}mez}, {Hibbard}, {Humphreys}, {Kent}, {Kepley}, {Kunneriath}, {Lipnicky}, {Loomis}, {Mason}, {Masters}, {Maud}, {Muders}, {Sabater}, {Sugimoto}, {Sz{\H{u}}cs}, {Vasiliev}, {Videla}, {Villard}, {Williams}, {Xue}, \& {Yoon}}]{Hunter+2023}
{Hunter}, T.~R., {Indebetouw}, R., {Brogan}, C.~L., {et~al.} 2023, \pasp, 135, 074501

\bibitem[{{Ito} {et~al.}(2025){Ito}, {Valentino}, {Brammer}, {Hamadouche}, {Whitaker}, {Kokorev}, {Zhu}, {Kakimoto}, {Wu}, {Antwi-Danso}, {Baker}, {Ceverino}, {Faisst}, {Farcy}, {Fujimoto}, {Gallazzi}, {Gillman}, {Gottumukkala}, {Heintz}, {Hirschmann}, {Jespersen}, {Kubo}, {Lee}, {Magdis}, {Onodera}, {Shimakawa}, {Tanaka}, {Toft}, \& {Weaver}}]{ito_2025_deepdive}
{Ito}, K., {Valentino}, F., {Brammer}, G., {et~al.} 2025, arXiv e-prints, arXiv:2506.22642

\bibitem[{{Izumi} {et~al.}(2019){Izumi}, {Onoue}, {Matsuoka}, {Nagao}, {Strauss}, {Imanishi}, {Kashikawa}, {Fujimoto}, {Kohno}, {Toba}, {Umehata}, {Goto}, {Ueda}, {Shirakata}, {Silverman}, {Greene}, {Harikane}, {Hashimoto}, {Ikarashi}, {Iono}, {Iwasawa}, {Lee}, {Minezaki}, {Nakanishi}, {Tamura}, {Tang}, \& {Taniguchi}}]{Izumi+2019}
{Izumi}, T., {Onoue}, M., {Matsuoka}, Y., {et~al.} 2019, \pasj, 71, 111

\bibitem[{{Izumi} {et~al.}(2018){Izumi}, {Onoue}, {Shirakata}, {Nagao}, {Kohno}, {Matsuoka}, {Imanishi}, {Strauss}, {Kashikawa}, \& {Schulze}}]{Izumi+2018}
{Izumi}, T., {Onoue}, M., {Shirakata}, H., {et~al.} 2018, \pasj, 70, 36

\bibitem[{{Ji} {et~al.}(2026){Ji}, {Williams}, {Behroozi}, {Weibel}, {Jespersen}, {Oesch}, {Bezanson}, {Whitaker}, {Greene}, {Brammer}, {Dayal}, {Labb{\'e}}, {Manning}, {Rinaldi}, {Xiao}, \& {Zhang}}]{ji_2026_number_density_qg}
{Ji}, Z., {Williams}, C.~C., {Behroozi}, P., {et~al.} 2026, arXiv e-prints, arXiv:2604.05022

\bibitem[{{Ji} {et~al.}(2024){Ji}, {Williams}, {Rieke}, {Lyu}, {Alberts}, {Sun}, {Helton}, {Rieke}, {Shivaei}, {D'Eugenio}, {Tacchella}, {Robertson}, {Zhu}, {Maiolino}, {Bunker}, {Sun}, \& {Willmer}}]{ji_2024}
{Ji}, Z., {Williams}, C.~C., {Rieke}, G.~H., {et~al.} 2024, arXiv:2409.17233

\bibitem[{{Johnson} {et~al.}(2021){Johnson}, {Leja}, {Conroy}, \& {Speagle}}]{johnson_2021_prospector}
{Johnson}, B.~D., {Leja}, J., {Conroy}, C., \& {Speagle}, J.~S. 2021, \apjs, 254, 22

\bibitem[{{Kennicutt} \& {Evans}(2012)}]{kennicutt-evans_2012}
{Kennicutt}, R.~C. \& {Evans}, N.~J. 2012, \araa, 50, 531

\bibitem[{{Kimmig} {et~al.}(2025){Kimmig}, {Remus}, {Seidel}, {Valenzuela}, {Dolag}, \& {Burkert}}]{kimmig_2025}
{Kimmig}, L.~C., {Remus}, R.-S., {Seidel}, B., {et~al.} 2025, \apj, 979, 15

\bibitem[{{Kriek} \& {Conroy}(2013)}]{kriek_2013}
{Kriek}, M. \& {Conroy}, C. 2013, \apjl, 775, L16

\bibitem[{{Kubo} {et~al.}(2021){Kubo}, {Umehata}, {Matsuda}, {Kajisawa}, {Steidel}, {Yamada}, {Tanaka}, {Hatsukade}, {Tamura}, {Nakanishi}, {Kohno}, {Lee}, \& {Matsuda}}]{kubo_2021}
{Kubo}, M., {Umehata}, H., {Matsuda}, Y., {et~al.} 2021, \apj, 919, 6

\bibitem[{{Kubo} {et~al.}(2022){Kubo}, {Umehata}, {Matsuda}, {Kajisawa}, {Steidel}, {Yamada}, {Tanaka}, {Hatsukade}, {Tamura}, {Nakanishi}, {Kohno}, {Lee}, {Matsuda}, {Ao}, {Nagao}, \& {Yun}}]{kubo_2022}
{Kubo}, M., {Umehata}, H., {Matsuda}, Y., {et~al.} 2022, \apj, 935, 89

\bibitem[{{Kubo} {et~al.}(2016){Kubo}, {Yamada}, {Ichikawa}, {Kajisawa}, {Matsuda}, {Tanaka}, \& {Umehata}}]{kubo_2016}
{Kubo}, M., {Yamada}, T., {Ichikawa}, T., {et~al.} 2016, \mnras, 455, 3333

\bibitem[{{Kurinchi-Vendhan} {et~al.}(2024){Kurinchi-Vendhan}, {Farcy}, {Hirschmann}, \& {Valentino}}]{kurinchi-vendhan_2024}
{Kurinchi-Vendhan}, S., {Farcy}, M., {Hirschmann}, M., \& {Valentino}, F. 2024, \mnras, 534, 3974

\bibitem[{{Lagos} {et~al.}(2024){Lagos}, {Bravo}, {Tobar}, {Obreschkow}, {Power}, {Robotham}, {Proctor}, {Hansen}, {Chandro-G{\'o}mez}, \& {Carrivick}}]{lagos_2024}
{Lagos}, C. d.~P., {Bravo}, M., {Tobar}, R., {et~al.} 2024, \mnras, 531, 3551

\bibitem[{{Lagos} {et~al.}(2025){Lagos}, {Valentino}, {Wright}, {de Graaff}, {Glazebrook}, {De Lucia}, {Robotham}, {Nanayakkara}, {Chandro-Gomez}, {Bravo}, {Baugh}, {Harborne}, {Hirschmann}, {Fontanot}, {Xie}, \& {Chittenden}}]{lagos_2025}
{Lagos}, C. d.~P., {Valentino}, F., {Wright}, R.~J., {et~al.} 2025, \mnras, 536, 2324

\bibitem[{{Le F{\`e}vre} {et~al.}(2020){Le F{\`e}vre}, {B{\'e}thermin}, {Faisst}, {Jones}, {Capak}, {Cassata}, {Silverman}, {Schaerer}, {Yan}, {Amorin}, {Bardelli}, {Boquien}, {Cimatti}, {Dessauges-Zavadsky}, {Giavalisco}, {Hathi}, {Fudamoto}, {Fujimoto}, {Ginolfi}, {Gruppioni}, {Hemmati}, {Ibar}, {Koekemoer}, {Khusanova}, {Lagache}, {Lemaux}, {Loiacono}, {Maiolino}, {Mancini}, {Narayanan}, {Morselli}, {M{\'e}ndez-Hern{\`a}ndez}, {Oesch}, {Pozzi}, {Romano}, {Riechers}, {Scoville}, {Talia}, {Tasca}, {Thomas}, {Toft}, {Vallini}, {Vergani}, {Walter}, {Zamorani}, \& {Zucca}}]{lefevre_2020}
{Le F{\`e}vre}, O., {B{\'e}thermin}, M., {Faisst}, A., {et~al.} 2020, \aap, 643, A1

\bibitem[{{Lee} {et~al.}(2024){Lee}, {Steidel}, {Brammer}, {F{\"o}rster-Schreiber}, {Renzini}, {Liu}, {Herrera-Camus}, {Naab}, {Price}, {{\"U}bler}, {Arriagada-Neira}, \& {Magdis}}]{Lee+2024}
{Lee}, M.~M., {Steidel}, C.~C., {Brammer}, G., {et~al.} 2024, \mnras, 527, 9529

\bibitem[{{Leja} {et~al.}(2019{\natexlab{a}}){Leja}, {Carnall}, {Johnson}, {Conroy}, \& {Speagle}}]{leja_2019_sfh}
{Leja}, J., {Carnall}, A.~C., {Johnson}, B.~D., {Conroy}, C., \& {Speagle}, J.~S. 2019{\natexlab{a}}, \apj, 876, 3

\bibitem[{{Leja} {et~al.}(2019{\natexlab{b}}){Leja}, {Johnson}, {Conroy}, {van Dokkum}, {Speagle}, {Brammer}, {Momcheva}, {Skelton}, {Whitaker}, {Franx}, \& {Nelson}}]{leja_2019_older_quiescent_universe}
{Leja}, J., {Johnson}, B.~D., {Conroy}, C., {et~al.} 2019{\natexlab{b}}, \apj, 877, 140

\bibitem[{{Liu} {et~al.}(2019){Liu}, {Schinnerer}, {Groves}, {Magnelli}, {Lang}, {Leslie}, {Jim{\'e}nez-Andrade}, {Riechers}, {Popping}, {Magdis}, {Daddi}, {Sargent}, {Gao}, {Fudamoto}, {Oesch}, \& {Bertoldi}}]{liu_2019}
{Liu}, D., {Schinnerer}, E., {Groves}, B., {et~al.} 2019, \apj, 887, 235

\bibitem[{{Lorenzon} {et~al.}(2025){Lorenzon}, {Donevski}, {Man}, {Romano}, {Whitaker}, {Belli}, {Liu}, {Lee}, {Narayanan}, {Long}, {Shivaei}, {Nanni}, {Lisiecki}, {Sawant}, {Rodighiero}, {Damjanov}, {Junais}, {Dav{\'e}}, {Pappalardo}, {Lovell}, \& {Hamed}}]{lorenzon_2025}
{Lorenzon}, G., {Donevski}, D., {Man}, A.~W.~S., {et~al.} 2025, \apjl, 995, L63

\bibitem[{{Madden} {et~al.}(2020){Madden}, {Cormier}, {Hony}, {Lebouteiller}, {Abel}, {Galametz}, {De Looze}, {Chevance}, {Polles}, {Lee}, {Galliano}, {Lambert-Huyghe}, {Hu}, \& {Ramambason}}]{madden_2020}
{Madden}, S.~C., {Cormier}, D., {Hony}, S., {et~al.} 2020, \aap, 643, A141

\bibitem[{{Magdis} {et~al.}(2021){Magdis}, {Gobat}, {Valentino}, {Daddi}, {Zanella}, {Kokorev}, {Toft}, {Jin}, \& {Whitaker}}]{magdis_2021}
{Magdis}, G.~E., {Gobat}, R., {Valentino}, F., {et~al.} 2021, \aap, 647, A33

\bibitem[{{Man} \& {Belli}(2018)}]{man_2018}
{Man}, A. \& {Belli}, S. 2018, Nature Astronomy, 2, 695

\bibitem[{{Mart{\'\i}-Vidal} {et~al.}(2014){Mart{\'\i}-Vidal}, {Vlemmings}, {Muller}, \& {Casey}}]{Marti-Vidal+2014}
{Mart{\'\i}-Vidal}, I., {Vlemmings}, W.~H.~T., {Muller}, S., \& {Casey}, S. 2014, \aap, 563, A136

\bibitem[{{McMullin} {et~al.}(2007){McMullin}, {Waters}, {Schiebel}, {Young}, \& {Golap}}]{McMullin+2007}
{McMullin}, J.~P., {Waters}, B., {Schiebel}, D., {Young}, W., \& {Golap}, K. 2007, in Astronomical Data Analysis Software and Systems XVI, Vol. 376, 127

\bibitem[{{Merlin} {et~al.}(2025){Merlin}, {Fortuni}, {Calabr{\'o}}, {Castellano}, {Santini}, {Fontana}, {Kimmig}, {Shankar}, {Napolitano}, {Koekemoer}, {Lucas}, {Pacucci}, {Cooper}, {Hirschmann}, {P{\'e}rez-Gonz{\'a}lez}, {Barro}, {Dickinson}, {Gandolfi}, {Paris}, {Grogin}, \& {Wang}}]{merlin_2025}
{Merlin}, E., {Fortuni}, F., {Calabr{\'o}}, A., {et~al.} 2025, The Open Journal of Astrophysics, 8, E170

\bibitem[{{Micha{\l}owski} {et~al.}(2024){Micha{\l}owski}, {Gall}, {Hjorth}, {Frayer}, {Tsai}, {Rowlands}, {Takeuchi}, {Le{\'s}niewska}, {Behrendt}, {Bourne}, {Hughes}, {Koprowski}, {Nadolny}, {Ryzhov}, {Solar}, {Spring}, {Zavala}, \& {Bartczak}}]{michalowski_2024}
{Micha{\l}owski}, M.~J., {Gall}, C., {Hjorth}, J., {et~al.} 2024, \apj, 964, 129

\bibitem[{{Micha{\l}owski} {et~al.}(2019){Micha{\l}owski}, {Hjorth}, {Gall}, {Frayer}, {Tsai}, {Hirashita}, {Rowlands}, {Takeuchi}, {Le{\'s}niewska}, {Behrendt}, {Bourne}, {Hughes}, {Spring}, {Zavala}, \& {Bartczak}}]{michalowski_2019}
{Micha{\l}owski}, M.~J., {Hjorth}, J., {Gall}, C., {et~al.} 2019, \aap, 632, A43

\bibitem[{{Nanayakkara} {et~al.}(2024){Nanayakkara}, {Glazebrook}, {Jacobs}, {Kawinwanichakij}, {Schreiber}, {Brammer}, {Esdaile}, {Kacprzak}, {Labbe}, {Lagos}, {Marchesini}, {Marsan}, {Oesch}, {Papovich}, {Remus}, \& {Tran}}]{nanayakkara_2024}
{Nanayakkara}, T., {Glazebrook}, K., {Jacobs}, C., {et~al.} 2024, Scientific Reports, 14, 3724

\bibitem[{{Newman} {et~al.}(2018){Newman}, {Belli}, {Ellis}, \& {Patel}}]{newman_2018}
{Newman}, A.~B., {Belli}, S., {Ellis}, R.~S., \& {Patel}, S.~G. 2018, \apj, 862, 125

\bibitem[{{Newman} {et~al.}(2025){Newman}, {Gu}, {Belli}, {Ellis}, {Gangula}, {Greene}, {Walsh}, {Suyu}, {Ertl}, {Caminha}, {Granata}, {Grillo}, {Schuldt}, {Barone}, {Bird}, {Glazebrook}, {Jafariyazani}, {Kriek}, {Matthews}, {Morishita}, {Nanayakkara}, {Pierel}, {Acebr\textbackslash'on}, {Bergamini}, {Cha}, {Diego}, {Foo}, {Frye}, {Fudamoto}, {Jee}, {Kamieneski}, {Koekemoer}, {Meena}, {Nishida}, {Oguri}, {Rosati}, \& {Zitrin}}]{newman_2025}
{Newman}, A.~B., {Gu}, M., {Belli}, S., {et~al.} 2025, arXiv e-prints, arXiv:2503.17478

\bibitem[{{Park} {et~al.}(2025){Park}, {Conroy}, {Johnson}, {Leja}, {Dotter}, \& {Cargile}}]{park_2025_alphaMC}
{Park}, M., {Conroy}, C., {Johnson}, B.~D., {et~al.} 2025, \apj, 994, 165

\bibitem[{{P{\'e}rez-Gonz{\'a}lez} {et~al.}(2025){P{\'e}rez-Gonz{\'a}lez}, {D'Eugenio}, {Rodr{\'\i}guez del Pino}, {Perna}, {{\"U}bler}, {Maiolino}, {Arribas}, {Cresci}, {Lamperti}, {Bunker}, {Carniani}, {Charlot}, {Willott}, {B{\"o}ker}, {Parlanti}, {Scholtz}, {Venturi}, {Barro}, {Costantin}, {Mart{\'\i}n-Navarro}, {Dunlop}, \& {Magee}}]{perez-gonzalez_2025}
{P{\'e}rez-Gonz{\'a}lez}, P.~G., {D'Eugenio}, F., {Rodr{\'\i}guez del Pino}, B., {et~al.} 2025, Nature Astronomy, 9, 1240

\bibitem[{{Popping} \& {P{\'e}roux}(2022)}]{popping_2022}
{Popping}, G. \& {P{\'e}roux}, C. 2022, \mnras, 513, 1531

\bibitem[{{Popping} {et~al.}(2023){Popping}, {Shivaei}, {Sanders}, {Jones}, {Pope}, {Reddy}, {Shapley}, {Coil}, \& {Kriek}}]{popping_2023}
{Popping}, G., {Shivaei}, I., {Sanders}, R.~L., {et~al.} 2023, \aap, 670, A138

\bibitem[{{Remus} \& {Kimmig}(2025)}]{remus_2025_magneticum}
{Remus}, R.-S. \& {Kimmig}, L.~C. 2025, \apj, 982, 30

\bibitem[{{Rowland} {et~al.}(2025){Rowland}, {Heintz}, {Algera}, {Stefanon}, {Hodge}, {Bouwens}, {Aravena}, {da Cunha}, {Dayal}, {Ferrara}, {Fisher}, {Gonz{\'a}lez}, {Inami}, {Komarova}, {de Looze}, {Nanayakkara}, {Ormerod}, {Pallottini}, {Pollock}, {Smit}, {van der Werf}, \& {Witstok}}]{rowland_2025}
{Rowland}, L.~E., {Heintz}, K.~E., {Algera}, H., {et~al.} 2025, arXiv e-prints, arXiv:2510.11351

\bibitem[{{Rowlands} {et~al.}(2012){Rowlands}, {Dunne}, {Maddox}, {Bourne}, {Gomez}, {Kaviraj}, {Bamford}, {Brough}, {Charlot}, {da Cunha}, {Driver}, {Eales}, {Hopkins}, {Kelvin}, {Nichol}, {Sansom}, {Sharp}, {Smith}, {Temi}, {van der Werf}, {Baes}, {Cava}, {Cooray}, {Croom}, {Dariush}, {de Zotti}, {Dye}, {Fritz}, {Hopwood}, {Ibar}, {Ivison}, {Liske}, {Loveday}, {Madore}, {Norberg}, {Popescu}, {Rigby}, {Robotham}, {Rodighiero}, {Seibert}, \& {Tuffs}}]{rowlands_2012}
{Rowlands}, K., {Dunne}, L., {Maddox}, S., {et~al.} 2012, \mnras, 419, 2545

\bibitem[{{Rowlands} {et~al.}(2015){Rowlands}, {Wild}, {Nesvadba}, {Sibthorpe}, {Mortier}, {Lehnert}, \& {da Cunha}}]{rowlands_2015}
{Rowlands}, K., {Wild}, V., {Nesvadba}, N., {et~al.} 2015, \mnras, 448, 258

\bibitem[{{S{\'a}nchez-Bl{\'a}zquez} {et~al.}(2006){S{\'a}nchez-Bl{\'a}zquez}, {Peletier}, {Jim{\'e}nez-Vicente}, {Cardiel}, {Cenarro}, {Falc{\'o}n-Barroso}, {Gorgas}, {Selam}, \& {Vazdekis}}]{sanchez-blazquez_2006}
{S{\'a}nchez-Bl{\'a}zquez}, P., {Peletier}, R.~F., {Jim{\'e}nez-Vicente}, J., {et~al.} 2006, \mnras, 371, 703

\bibitem[{{Schaerer} {et~al.}(2020){Schaerer}, {Ginolfi}, {B{\'e}thermin}, {Fudamoto}, {Oesch}, {Le F{\`e}vre}, {Faisst}, {Capak}, {Cassata}, {Silverman}, {Yan}, {Jones}, {Amorin}, {Bardelli}, {Boquien}, {Cimatti}, {Dessauges-Zavadsky}, {Giavalisco}, {Hathi}, {Fujimoto}, {Ibar}, {Koekemoer}, {Lagache}, {Lemaux}, {Loiacono}, {Maiolino}, {Narayanan}, {Morselli}, {M{\'e}ndez-Hern{\`a}ndez}, {Pozzi}, {Riechers}, {Talia}, {Toft}, {Vallini}, {Vergani}, {Zamorani}, \& {Zucca}}]{schaerer_2020}
{Schaerer}, D., {Ginolfi}, M., {B{\'e}thermin}, M., {et~al.} 2020, \aap, 643, A3

\bibitem[{{Setton} {et~al.}(2025){Setton}, {Spilker}, {Bezanson}, {Suess}, {Greene}, {Goulding}, {Cenci}, {D'Onofrio}, {Feldmann}, {Kriek}, {Kumar}, {Luo}, {Narayanan}, {Verrico}, \& {Zhu}}]{setton_2025}
{Setton}, D.~J., {Spilker}, J.~S., {Bezanson}, R., {et~al.} 2025, \aj, 170, 351

\bibitem[{{Smercina} {et~al.}(2018){Smercina}, {Smith}, {Dale}, {French}, {Croxall}, {Zhukovska}, {Togi}, {Bell}, {Crocker}, {Draine}, {Jarrett}, {Tremonti}, {Yang}, \& {Zabludoff}}]{smercina_2018}
{Smercina}, A., {Smith}, J.~D.~T., {Dale}, D.~A., {et~al.} 2018, \apj, 855, 51

\bibitem[{{Smith} {et~al.}(2012){Smith}, {Gomez}, {Eales}, {Ciesla}, {Boselli}, {Cortese}, {Bendo}, {Baes}, {Bianchi}, {Clemens}, {Clements}, {Cooray}, {Davies}, {De Looze}, {di Serego Alighieri}, {Fritz}, {Gavazzi}, {Gear}, {Madden}, {Mentuch}, {Panuzzo}, {Pohlen}, {Spinoglio}, {Verstappen}, {Vlahakis}, {Wilson}, \& {Xilouris}}]{smith_2012}
{Smith}, M.~W.~L., {Gomez}, H.~L., {Eales}, S.~A., {et~al.} 2012, \apj, 748, 123

\bibitem[{{Solomon} {et~al.}(1997){Solomon}, {Downes}, {Radford}, \& {Barrett}}]{Solomon+1997}
{Solomon}, P.~M., {Downes}, D., {Radford}, S.~J.~E., \& {Barrett}, J.~W. 1997, \apj, 478, 144

\bibitem[{{Spilker} {et~al.}(2018){Spilker}, {Bezanson}, {Bari{\v{s}}i{\'c}}, {Bell}, {Lagos}, {Maseda}, {Muzzin}, {Pacifici}, {Sobral}, {Straatman}, {van der Wel}, {van Dokkum}, {Weiner}, {Whitaker}, {Williams}, \& {Wu}}]{spilker_2018}
{Spilker}, J., {Bezanson}, R., {Bari{\v{s}}i{\'c}}, I., {et~al.} 2018, \apj, 860, 103

\bibitem[{{Spilker} {et~al.}(2025){Spilker}, {Whitaker}, {Narayanan}, {Bezanson}, {Bodansky}, {D'Onofrio}, {Feldmann}, {Goulding}, {Greene}, {Kriek}, {Luo}, {Setton}, {Suess}, {van der Wel}, {Verrico}, {Williams}, {Woodrum}, \& {Wu}}]{spilker_2025}
{Spilker}, J.~S., {Whitaker}, K.~E., {Narayanan}, D., {et~al.} 2025, \apjl, 993, L40

\bibitem[{{Stevenson} {et~al.}(2026){Stevenson}, {Carnall}, {Leung}, {Taylor}, {Cullen}, {Dunlop}, {McLeod}, {McLure}, {Begley}, {Arellano-C{\'o}rdova}, {Barrufet}, {Bondestam}, {Donnan}, {Ellis}, {Grogin}, {Koekemoer}, {Liu}, {P{\'e}rez-Gonz{\'a}lez}, {Rowlands}, {Sanders}, {Scholte}, {Shapley}, {Skarbinski}, {Stanton}, \& {Wild}}]{stevenson_2026}
{Stevenson}, S.~D., {Carnall}, A.~C., {Leung}, H.-H., {et~al.} 2026, \mnras, 545, staf2087

\bibitem[{{Suess} {et~al.}(2017){Suess}, {Bezanson}, {Spilker}, {Kriek}, {Greene}, {Feldmann}, {Hunt}, \& {Narayanan}}]{suess_2017}
{Suess}, K.~A., {Bezanson}, R., {Spilker}, J.~S., {et~al.} 2017, \apjl, 846, L14

\bibitem[{{Suess} {et~al.}(2023){Suess}, {Williams}, {Robertson}, {Ji}, {Johnson}, {Nelson}, {Alberts}, {Hainline}, {D'Eugenio}, {{\"U}bler}, {Rieke}, {Rieke}, {Bunker}, {Carniani}, {Charlot}, {Eisenstein}, {Maiolino}, {Stark}, {Tacchella}, \& {Willott}}]{suess_2023}
{Suess}, K.~A., {Williams}, C.~C., {Robertson}, B., {et~al.} 2023, \apjl, 956, L42

\bibitem[{{Sun} {et~al.}(2026){Sun}, {Ji}, {D'Eugenio}, {Zhu}, {Rieke}, {Baker}, {Bunker}, {Carniani}, {Helton}, {Perna}, {P{\'e}rez-Gonz{\'a}lez}, {Rinaldi}, {{\"U}bler}, \& {Willmer}}]{sun_2026_jades_outflows}
{Sun}, Y., {Ji}, Z., {D'Eugenio}, F., {et~al.} 2026, arXiv e-prints, arXiv:2604.18522

\bibitem[{{Suzuki} {et~al.}(2022){Suzuki}, {Glazebrook}, {Schreiber}, {Kodama}, {Kacprzak}, {Leiton}, {Nanayakkara}, {Oesch}, {Papovich}, {Spitler}, {Straatman}, {Tran}, \& {Wang}}]{suzuki_2022}
{Suzuki}, T.~L., {Glazebrook}, K., {Schreiber}, C., {et~al.} 2022, \apj, 936, 61

\bibitem[{{Taylor} {et~al.}(2026){Taylor}, {Carnall}, {Maltby}, {Almaini}, {Leung}, {Stevenson}, {Negri}, {Cullen}, {Wild}, {McLure}, {Shapley}, {Arellano-C{\'o}rdova}, {Begley}, {Bondestam}, {de Lisle}, {Donnan}, {Dunlop}, {Ellis}, {Hewitt}, {Koekemoer}, {Frey Liu}, {McLeod}, {Rowlands}, {Sanders}, {Scholte}, {Skarbinski}, \& {Stanton}}]{taylor_2026}
{Taylor}, E., {Carnall}, A.~C., {Maltby}, D., {et~al.} 2026, arXiv e-prints, arXiv:2601.02269

\bibitem[{Tazzari(2017)}]{Tazzari2017}
Tazzari, M. 2017, mtazzari/uvplot

\bibitem[{{Umehata} {et~al.}(2025){Umehata}, {Kubo}, \& {Nakanishi}}]{umehata_2025}
{Umehata}, H., {Kubo}, M., \& {Nakanishi}, K. 2025, \apjl, 985, L8

\bibitem[{{Umehata} {et~al.}(2026){Umehata}, {Kubo}, {Smail}, {Lehmer}, {Monson}, {Nakanishi}, \& {Matsuda}}]{umehata_2026}
{Umehata}, H., {Kubo}, M., {Smail}, I., {et~al.} 2026, \apj, 997, 79

\bibitem[{{Utomo} {et~al.}(2014){Utomo}, {Kriek}, {Labb{\'e}}, {Conroy}, \& {Fumagalli}}]{utomo_2014}
{Utomo}, D., {Kriek}, M., {Labb{\'e}}, I., {Conroy}, C., \& {Fumagalli}, M. 2014, \apjl, 783, L30

\bibitem[{{Valentino} {et~al.}(2023){Valentino}, {Brammer}, {Gould}, {Kokorev}, {Fujimoto}, {Jespersen}, {Vijayan}, {Weaver}, {Ito}, {Tanaka}, {Ilbert}, {Magdis}, {Whitaker}, {Faisst}, {Gallazzi}, {Gillman}, {Gim{\'e}nez-Arteaga}, {G{\'o}mez-Guijarro}, {Kubo}, {Heintz}, {Hirschmann}, {Oesch}, {Onodera}, {Rizzo}, {Lee}, {Strait}, \& {Toft}}]{valentino_2023}
{Valentino}, F., {Brammer}, G., {Gould}, K. M.~L., {et~al.} 2023, \apj, 947, 20

\bibitem[{{Valentino} {et~al.}(2025){Valentino}, {Heintz}, {Brammer}, {Ito}, {Kokorev}, {Whitaker}, {Gallazzi}, {de Graaff}, {Weibel}, {Frye}, {Kamieneski}, {Jin}, {Ceverino}, {Faisst}, {Farcy}, {Fujimoto}, {Gillman}, {Gottumukkala}, {Hamadouche}, {Harrington}, {Hirschmann}, {Jespersen}, {Kakimoto}, {Kubo}, {Lagos}, {Lee}, {Magdis}, {Man}, {Onodera}, {Rizzo}, {Shimakawa}, {Setton}, {Tanaka}, {Toft}, {Wu}, \& {Zhu}}]{valentino_2025}
{Valentino}, F., {Heintz}, K.~E., {Brammer}, G., {et~al.} 2025, \aap, 699, A358

\bibitem[{{Vallini} {et~al.}(2015){Vallini}, {Gallerani}, {Ferrara}, {Pallottini}, \& {Yue}}]{vallini_2015}
{Vallini}, L., {Gallerani}, S., {Ferrara}, A., {Pallottini}, A., \& {Yue}, B. 2015, \apj, 813, 36

\bibitem[{{Vizgan} {et~al.}(2022{\natexlab{a}}){Vizgan}, {Greve}, {Olsen}, {Zanella}, {Narayanan}, {Dav{\`e}}, {Magdis}, {Popping}, {Valentino}, \& {Heintz}}]{vizgan_2022_cii_molecular}
{Vizgan}, D., {Greve}, T.~R., {Olsen}, K.~P., {et~al.} 2022{\natexlab{a}}, \apj, 929, 92

\bibitem[{{Vizgan} {et~al.}(2022{\natexlab{b}}){Vizgan}, {Heintz}, {Greve}, {Narayanan}, {Dav{\'e}}, {Olsen}, {Popping}, \& {Watson}}]{vizgan_2022_cii_neutral}
{Vizgan}, D., {Heintz}, K.~E., {Greve}, T.~R., {et~al.} 2022{\natexlab{b}}, \apjl, 939, L1

\bibitem[{{Wang} {et~al.}(2026){Wang}, {Champagne}, {Huang}, {Yang}, {Hennawi}, {Fan}, {Zhang}, {Costa}, {Decarli}, {Habouzit}, {Sun}, {Banados}, {Jin}, {Kakiichi}, {Meyer}, {Wu}, {Belladitta}, {Blecha}, {Bosman}, {Cai}, {Connor}, {Davies}, {Eilers}, {Haiman}, {Jun}, {Li}, {Li}, {Liu}, {Lupi}, {Lyu}, {Mazzucchelli}, {Onoue}, {Pudoka}, {Rojas-Ruiz}, {Schindler}, {Shen}, {Tee}, {Trakhtenbrot}, {Trebitsch}, {Vestergaard}, {Volonteri}, {Walter}, {Zhang}, \& {Zou}}]{wang_2026_aspire_survey_paper}
{Wang}, F., {Champagne}, J.~B., {Huang}, J., {et~al.} 2026, arXiv e-prints, arXiv:2602.04979

\bibitem[{{Weibel} {et~al.}(2025){Weibel}, {de Graaff}, {Setton}, {Miller}, {Oesch}, {Brammer}, {Lagos}, {Whitaker}, {Williams}, {Baggen}, {Bezanson}, {Boogaard}, {Cleri}, {Greene}, {Hirschmann}, {Hviding}, {Kuruvanthodi}, {Labb{\'e}}, {Leja}, {Maseda}, {Matthee}, {McConachie}, {Naidu}, {Roberts-Borsani}, {Schaerer}, {Suess}, {Valentino}, {van Dokkum}, \& {Wang}}]{weibel_2025}
{Weibel}, A., {de Graaff}, A., {Setton}, D.~J., {et~al.} 2025, \apj, 983, 11

\bibitem[{{Whitaker} \& {Bezanson}(2026)}]{whitaker-bezanson_2026}
{Whitaker}, K.~E. \& {Bezanson}, R. 2026, arXiv e-prints, arXiv:2606.12156

\bibitem[{{Whitaker} {et~al.}(2021{\natexlab{a}}){Whitaker}, {Narayanan}, {Williams}, {Li}, {Spilker}, {Dav{\'e}}, {Akhshik}, {Akins}, {Bezanson}, {Katz}, {Leja}, {Magdis}, {Mowla}, {Nelson}, {Pope}, {Privon}, {Toft}, \& {Valentino}}]{whitaker_2021}
{Whitaker}, K.~E., {Narayanan}, D., {Williams}, C.~C., {et~al.} 2021{\natexlab{a}}, \apjl, 922, L30

\bibitem[{{Whitaker} {et~al.}(2021{\natexlab{b}}){Whitaker}, {Williams}, {Mowla}, {Spilker}, {Toft}, {Narayanan}, {Pope}, {Magdis}, {van Dokkum}, {Akhshik}, {Bezanson}, {Brammer}, {Leja}, {Man}, {Nelson}, {Richard}, {Pacifici}, {Sharon}, \& {Valentino}}]{whitaker_2021_detection}
{Whitaker}, K.~E., {Williams}, C.~C., {Mowla}, L., {et~al.} 2021{\natexlab{b}}, \nat, 597, 485

\bibitem[{{Williams} {et~al.}(2021){Williams}, {Spilker}, {Whitaker}, {Dav{\'e}}, {Woodrum}, {Brammer}, {Bezanson}, {Narayanan}, \& {Weiner}}]{williams_2021_co_quiescent}
{Williams}, C.~C., {Spilker}, J.~S., {Whitaker}, K.~E., {et~al.} 2021, \apj, 908, 54

\bibitem[{{Witstok} {et~al.}(2023){Witstok}, {Jones}, {Maiolino}, {Smit}, \& {Schneider}}]{witstok_2023}
{Witstok}, J., {Jones}, G.~C., {Maiolino}, R., {Smit}, R., \& {Schneider}, R. 2023, \mnras, 523, 3119

\bibitem[{{Wolfire} {et~al.}(2022){Wolfire}, {Vallini}, \& {Chevance}}]{wolfire_2022}
{Wolfire}, M.~G., {Vallini}, L., \& {Chevance}, M. 2022, \araa, 60, 247

\bibitem[{{Woodrum} {et~al.}(2022){Woodrum}, {Williams}, {Rieke}, {Leja}, {Johnson}, {Bezanson}, {Kennicutt}, {Spilker}, \& {Tacchella}}]{woodrum_2022}
{Woodrum}, C., {Williams}, C.~C., {Rieke}, M., {et~al.} 2022, \apj, 940, 39

\bibitem[{{Wu}(2025)}]{wu_2025}
{Wu}, P.-F. 2025, \apj, 978, 131

\bibitem[{{Xie} {et~al.}(2024){Xie}, {De Lucia}, {Fontanot}, {Hirschmann}, {Bah{\'e}}, {Balogh}, {Muzzin}, {Vulcani}, {Baxter}, {Forrest}, {Wilson}, {Rudnick}, {Cooper}, \& {Rescigno}}]{xie_2024}
{Xie}, L., {De Lucia}, G., {Fontanot}, F., {et~al.} 2024, \apjl, 966, L2

\bibitem[{{Zanella} {et~al.}(2026){Zanella}, {Belli}, {Valentino}, \& {Bolamperti}}]{zanella_2026}
{Zanella}, A., {Belli}, S., {Valentino}, F.~M., \& {Bolamperti}, A. 2026, \aap, 709, A14

\bibitem[{{Zanella} {et~al.}(2018){Zanella}, {Daddi}, {Magdis}, {Diaz Santos}, {Cormier}, {Liu}, {Cibinel}, {Gobat}, {Dickinson}, {Sargent}, {Popping}, {Madden}, {Bethermin}, {Hughes}, {Valentino}, {Rujopakarn}, {Pannella}, {Bournaud}, {Walter}, {Wang}, {Elbaz}, \& {Coogan}}]{zanella_2018}
{Zanella}, A., {Daddi}, E., {Magdis}, G., {et~al.} 2018, \mnras, 481, 1976

\bibitem[{{Zanella} {et~al.}(2023){Zanella}, {Valentino}, {Gallazzi}, {Belli}, {Magdis}, \& {Bolamperti}}]{zanella_2023}
{Zanella}, A., {Valentino}, F., {Gallazzi}, A., {et~al.} 2023, \mnras, 524, 923

\bibitem[{{Zhang} {et~al.}(2026){Zhang}, {de Graaff}, {Setton}, {Price}, {Bezanson}, {del P. Lagos}, {Cutler}, {McConachie}, {Cleri}, {Cooper}, {Gottumukkala}, {Greene}, {Hirschmann}, {Khullar}, {Labbe}, {Leja}, {Maseda}, {Matthee}, {Miller}, {Nanayakkara}, {Suess}, {Wang}, {Whitaker}, \& {Williams}}]{zhang_2026_rubies_number_densities}
{Zhang}, Y., {de Graaff}, A., {Setton}, D.~J., {et~al.} 2026, \apj, 997, 252

\bibitem[{{Zhu} {et~al.}(2026){Zhu}, {Ito}, {Valentino}, {Hamadouche}, {Scarpe}, {Whitaker}, {Kakimoto}, {Baker}, {Gallazzi}, {Gillman}, {Gottumukkala}, {Jespersen}, {Lee}, {Man}, {Magdis}, {Onodera}, {Shimakawa}, {Vijayan}, \& {Wu}}]{zhu_2026_outflows}
{Zhu}, P., {Ito}, K., {Valentino}, F., {et~al.} 2026, arXiv e-prints, arXiv:2602.17767

\end{thebibliography}

\begin{table}
\caption{Physical properties of \rubies.}
\label{tab:physical_properties}
\centering
\begin{tabular}{lcc}
\toprule
\toprule
Quantity & \lowZ & \highZ \\
\midrule
$\log(M_\star/M_\odot)^\dagger$
    & $10.23^{+0.04}_{-0.04}$
    & $10.19^{+0.04}_{-0.04}$ \\
$A_V$ [mag]$^\dagger$
    & $0.31^{+0.08}_{-0.08}$
    & $0.25^{+0.09}_{-0.07}$ \\
$t_{50}$ [Gyr]$^\dagger$
    & $0.20^{+0.07}_{-0.02}$
    & $0.16^{+0.03}_{-0.02}$ \\
$t_{90}$ [Gyr]$^\dagger$
    & $0.12^{+0.01}_{-0.01}$
    & $0.07^{+0.01}_{-0.01}$ \\
$\log(Z/Z_\odot)$$^\dagger$
    & $-0.94^{+0.05}_{-0.04}$
    & $0.07^{+0.08}_{-0.11}$ \\
$\rm SFR_{10}$ [\myr]$^\dagger$
    & $0.64^{+0.83}_{-0.60}$
    & $1.08^{+1.55}_{-0.98}$ \\
$\rm SFR_{100}$ [\myr]$^\dagger$
    & $0.84^{+20.16}_{-0.78}$
    & $48.89^{+21.12}_{-13.04}$ \\
$\rm SFR_{Lines}$ [\myr]$^\dagger$& \multicolumn{2}{c}{$<6.6$}\\
\midrule
$\log(M_{\rm HI}/M_\odot)$ (\cii, H21) & $10.34^{+0.19}_{-0.17}$ & $9.46^{+0.18}_{-0.17}$ \\
$\log(M_{\rm mol}/M_\odot)$ (\cii, Z18)   & \multicolumn{2}{c}{$9.53^{+0.32}_{-0.31}$} \\
$\log(M_{\rm gas}/M_\odot)$ (dust, P23) & $<10.44$ & $<9.13$ \\
% Gas fractions
$f_{\rm HI}$ (H21) & $128.6^{+58.5}_{-51.8}\%$ & $18.6^{+8.0}_{-7.6}\%$ \\
$f_{\rm mol}$ (Z18) & $20.1^{+15.0}_{-14.5}\%$& $22.0^{+16.5}_{-15.9}\%$ \\
$f_{\rm gas}$ (dust, P23) & $<163\%$ & $<9\%$ \\
$\tau_{\rm dep}(M_{\rm HI},\,\rm SFR_{10})$ [Gyr]
    & $>8.77$
    & $>0.62$ \\
$\tau_{\rm dep}(M_{\rm HI},\,\rm SFR_{100})$ [Gyr]
    & $>0.36$
    & $>0.05$ \\
$\tau_{\rm dep}(M_{\rm mol},\,\rm SFR_{10})$ [Gyr]
    & $>1.37$
    & $>0.73$ \\
$\tau_{\rm dep}(M_{\rm mol},\,\rm SFR_{100})$ [Gyr]
    & $>0.06$
    & $>0.05$ \\
\midrule
\multicolumn{3}{c}{Extended \cii\ emission}  \\
\cmidrule(lr){1-3}
$\log(M_{\rm HI}/M_\odot)$ (\cii, H21) & $10.66^{+0.22}_{-0.20}$ & $9.82^{+0.20}_{-0.20}$ \\
$\log(M_{\rm mol}/M_\odot)$ (\cii, Z18)   & \multicolumn{2}{c}{$9.85^{+0.34}_{-0.33}$} \\
\bottomrule
\end{tabular}
\tablefoot{$^\dagger$ Properties derived from the spectrophotometric modeling in \cite{weibel_2025} and reported here for convenience. SFR$_{10}$ and SFR$_{100}$
are the star formation rates averaged over the last
10 and 100 Myr, respectively.
$t_{50}$ and $t_{90}$ are the lookback times by
which 50\% and 90\% of the stellar mass had
already formed. Gas mass estimates are obtained from \cii\ (\hi, H21: \citealt{heintz_2021}; molecular, Z18: \citealt{zanella_2018}) and dust (P23: \citealt{popping_2023}). The upper limits are at $3\sigma$.}
\end{table}

\end{document}